\documentclass{article}

\usepackage{amssymb,amsfonts,amsmath}
\usepackage{cite,enumerate,float}
\usepackage{color}
\usepackage{tikz}
\usetikzlibrary{arrows,snakes,backgrounds}

\def\be{\begin{eqnarray}}
\def\ee{\end{eqnarray}}
\def\nn{\nonumber}

\def\p{\partial}

\def\wcs{weak compositions}

\definecolor{red}{rgb}{1,0,0}
\definecolor{orange}{rgb}{1,0.5,0}
\definecolor{violet}{rgb}{0.7,0,1}

\hoffset= - 1.0in         

\begin{document}

\title{\vspace{1.5cm}\bf
Twisted Cherednik spectrum as a $q,t$-deformation
}

\author{
A. Mironov$^{b,c,d,}$\footnote{mironov@lpi.ru,mironov@itep.ru},
A. Morozov$^{a,c,d,}$\footnote{morozov@itep.ru},
A. Popolitov$^{a,c,d,}$\footnote{popolit@gmail.com}
}

\date{ }

\maketitle

\vspace{-6cm}

\begin{center}
  \hfill MIPT/TH-01/26\\
  \hfill FIAN/TD-01/26\\
  \hfill ITEP/TH-01/26\\
  \hfill IITP/TH-01/26
\end{center}

\vspace{4.5cm}

\begin{center}
$^a$ {\small {\it MIPT, Dolgoprudny, 141701, Russia}}\\
$^b$ {\small {\it Lebedev Physics Institute, Moscow 119991, Russia}}\\
$^c$ {\small {\it NRC ``Kurchatov Institute", 123182, Moscow, Russia}}\\
$^d$ {\small {\it Institute for Information Transmission Problems, Moscow 127994, Russia}}
\end{center}

\vspace{.1cm}

\begin{abstract}
The common eigenfunctions of the twisted Cherednik operators
can be first analyzed in the limit of $q\longrightarrow 1$.
Then, the polynomial eigenfunctions form a simple set
originating from the symmetric ground state of non-vanishing degree and excitations over it,
described by non-symmetric polynomials of higher degrees and enumerated by
weak compositions.
This pattern is inherited by the full spectrum at $q\neq 1$,
which can be considered as a deformation.
The whole story looks   like a   typical NP problem:
the Cherednik equations are difficult to solve,
but easy to check the solution once it is somehow found.
\end{abstract}

\bigskip

\newcommand\smallpar[1]{
  \noindent $\bullet$ \textbf{#1}
}


\section{Introduction}

Integrable theories play a central role in modern theoretical physics, since it has been recognized \cite{UFN2,UFN3}
that they capture the basic properties of exact integrals and thus of the non-perturbative physics.
A new chapter is now opened by the discovery \cite{MMMP1,MMMP2,MMP} of a whole system of commutative subalgebras associated with many-body integrable systems within $W_{1+\infty}$ \cite{Pope,FKRN,Awata,KR2}, affine Yangian \cite{Ts,Proch} and Ding-Iohara-Miki (DIM) \cite{DI,Miki} algebras, which implies coexistence of large sets of integrable systems,
a phenomenon which promises new insights and applications,
though it is still awaiting an adequate physical interpretation.
In fact, at the moment, even the formal part of the story is not fully understood,
in the sense that appropriate formulas are not found and studied for all eigenfunctions,
so that their structure and properties are not fully revealed.
This is non-trivial, because the problem has enhanced but not explicit symmetries,
which need to be treated in some adequate language, which still needs to be found.

In this paper, we continue our study of the twisted Cherednik integrable system \cite{MMPns}, which is given by a set of Hamiltonians that we used while constructing commutative subalgebras of the DIM algebra in the $N$-body representation. Integrable Hamiltonians $\hat {\mathfrak{C}}_i^{(a)}[\vec x]$ in question are provided by
the sophisticated finite-difference twisted Cherednik operators, see \cite[(61)]{MMPns}.
They act on functions of $N$ variables $x_i$, $i=1,\ldots,N$,  and
are labeled by an additional twist parameter $a$. This same parameter $a$ enumerates the integer DIM rays \cite{MMP} associated with commuting subalgebras of the DIM algebra, in accordance with the relation \cite{DIMDAHA} between the DIM (or Elliptic Hall \cite{K,BS,S,Feigin}) algebra and spherical DAHA \cite{Ch}.

Our goal in \cite{MMPns} was to study eigenfunctions of these Hamiltonians, and here we continue this. In fact, we are interested in the class of eigenfunctions that become polynomial at peculiar values of parameter $t=q^{-m}$, $m\in\mathbb{N}$, and we study these polynomials. Hence, hereafter all the quantities are treated as depending on the parameter $q$ and the natural number $m$.

Eigenfunctions have \cite{MMPns} a hierarchical structure of a Verma module type growing from ``the ground state'' eigenfunction $\Omega_N^{(a,m)}[\vec x]$. As any ground state, it is a symmetric function of its variables.
The simplest way to obtain $\Omega_N^{(a,m)}$ is to note that it is proportional to a peculiar \cite{MMPns} twisted \cite{ChE} multivariable Baker-Akhiezer function \cite{Cha}. Indeed, the eigenfunctions of the commuting Hamiltonians associated with integer rays of the DIM algebra are just twisted multivariable Baker-Akhiezer functions \cite{ChF,MMPCha}. On the other hand, these Hamiltonians coincide with the power sums of the Cherednik twisted Hamiltonians $\hat {\mathfrak{C}}_i^{(a)}[\vec x]$ when acting on symmetric functions \cite{MMP}. Hence, any symmetric twisted Baker-Akhiezer function is simultaneously an eigenfunction of $\hat {\mathfrak{C}}_i^{(a)}[\vec x]$, it is just the ground state $\Omega_N^{(a,m)}$.

Now note that any twisted Baker-Akhiezer function can be obtained by solving the periodicity Chalykh equations \cite{Cha}, in the ``$a$-twisted'' parameters $X_i:=x_i^{1\over a}$, $Q:=q^{1\over a}$ looking as
\be
\Omega^{(a,m)}_{N}[\vec X]\Big|_{X_k=Q^jX,X_l=\varepsilon X}-\varepsilon^j\cdot \Omega^{(a,m)}_{N}[\vec X]\Big|_{X_k=X,X_l=\varepsilon Q^jX}
\sim (\varepsilon^a-1) \ \ \ \ \ \ \ \forall \ 1\leq j \leq m,\ \ \ \varepsilon^a=1
\label{peri}
\ee
at any $1\le k,l\le N$, and $\Omega^{(a,m)}_{N}[\vec X]$ being a peculiar symmetric twisted Baker-Akhiezer function (with parameters subject to the symmetricity condition\footnote{In the notation of \cite[Eq.(41)]{MMPCha}, they are equal to $\lambda_i=(i-1)m+m{(a-1)(N-1)\over 2}$.}), \cite{MMPns} have a polynomial form
\be \label{om}
\Omega^{(a,m)}_{N}[\vec X]=\left(\prod_{i=1}^NX_i^{(a-1)(N-i)m+l}\right)\cdot\sum_{K}c(a,m,k_{ij})
\prod_{i>j}\left({X_j\over X_i}\right)^{k_{ij}}
\ee
where the sum runs over the polytope $K:=\{k_{ij},\ i>j:\ 0\le k_{ij}\le am\}$, and $l$ has to be an arbitrary non-negative integer in order to have a polynomial. At the ground state, $l=0$ (see also (\ref{ereduct}), (\ref{Zreduct})).
As soon as $\Omega^{(a,m)}_{N}[\vec X]$ is a symmetric function of variables, it is sufficient to impose the periodicity conditions (\ref{peri}) only on the first two variables $X_1$ and $X_2$, and additionally to require symmetricity under all permutations.

Note that extending $\Omega^{(a,m)}_{N}[\vec X]$ to $t \ne q^{-m}$, though possible (see, e.g., \cite[Eq.(76)]{MMPns}) gives rise to non-polynomial objects depending on $t$ in a complicated way.

Despite to use the periodicity conditions is technically much simpler than looking for (twisted) Cherednik eigenfunctions,
the actual limits are still severe: one can hardly move to high values of $a$ and $m$.
Even for $N=3$, the calculations look practical only for a very limited region of parameters.

The question is how to get rid of equations at all and
directly describe the {\it answers} for $\Omega_{N}^{(a,m)}$, which are complicated
symmetric polynomials of $X_i$ of the degree ${\rm deg}_{N,a,m} = \frac{N(N-1)}{2}\cdot (a-1)\cdot m$ (see (\ref{om})).
Note that polynomial solutions of all higher degrees exist, but they are no longer symmetric,
and describe {\it excitations} over the symmetric background state $\Omega_{N}^{(a,m)}$.

The coefficients $c(a,m,k_{ij})$ of the symmetric solution (\ref{om}) are rather complicated functions of $q$. At the same time, the full expression is a $q$-deformation
of an extremely simple $q\to 1$ expression $\omega_N^{(a,m)} = \left(\prod_{i<j}^N \frac{X_i^{a}-X_j^{a}}{X_i-X_j}\right)^m$,
where the sophisticated ($q$-independent) coefficients of expansions to a polynomial similar to (\ref{om})
are immediate.
However, the problem is to find a proper deformation procedure, i.e. to find proper multiplicatively deformed basic polynomials for $\omega_N^{(a,m)}$ to expand in. This means that one needs an expansion for $\omega_N^{(a,m)}$ in such a form that it is sufficient just to make a replacement of all numbers to $q$-numbers in order to obtain the full $\Omega_N^{(a,m)}$, without need of any additional deformations.  Ideally, the recipe should be analogous to the
transition from the Schubert to $q$-Schubert polynomials \cite{paper:FGP},
where a the suitable basis of nested $e_{\vec{k}}$-polynomials, quantization
is completely algorithmic and amounts to substitution $e_{\vec{k}} \rightarrow E_{\vec{k}}$.

It is easy to see that the individual monomials $X^{\vec n}:=\prod_i X_{i}^{n_i}$
are not such ``proper'' expressions: the coefficients of them are not simply deformed,
neither as a total, nor as products of simple factors.
In order to demonstrate this, consider the first non-trivial sequence of such coefficients
$\{1,3,6,7,6,3,1\}$ appearing in front of $X_1^iX_2^{6-i}$ in $\omega_2^{(3,3)}$.
It turns out that the right decomposition is
\be
\Big\{\underline{\underline{1,3,6}},7,\boxed{6},3,1\Big\}
=\underbrace{\Big\{\underline{\underline{1,3,6}},10,\boxed{15},21,28\Big\}}_{A_i}-
3\cdot\underbrace{\Big\{0,0,0,\underline{\underline{1,\boxed{3},6,10}}\Big\}}_{A_{i-3}}+3\cdot \underbrace{\Big\{0,0,0,0,\boxed{0},0,\underline{\underline{1}}\Big\}}_{A_{i-6}}
\label{exadeco0}
\ee
with one and the same the sequence $A_i:=\frac{i(i+1)}{2}$ and $A_{-i}=0$ at $i\in\mathbb{Z}_{\ge 0}$, shifted by multiples of $a=3$.
Elements of the sequence are now deformed in an obvious way:
$\frac{i(i+1)}{2}\longrightarrow\frac{[i][i+1]}{[2]}$, but coefficients $3$ are further decomposed
$3\longrightarrow 2+1 \longrightarrow [2]\oplus [1]$.
All the emerging terms are after that multiplicatively deformed: multiplied by various powers of $q$,
which also need to be specified.
Indeed, in this example (\ref{exadeco0}), one needs the $q$-deformation
of equation like $\frac{4\cdot 3}{2}=\frac{6\cdot 5}{2} - (2+1)\cdot \frac{3\cdot 2}{2}$ associated with the boxed numbers,
which requires introducing additional factors $C_j=q^{\alpha_j}$:
\be
\frac{ \frac{q^4-1}{q-1}\cdot \frac{q^3-1}{q-1} }{\frac{q^2-1}{q-1}}  =
C_1\frac{  \frac{q^6-1}{q-1}\cdot \frac{q^5-1}{q-1} }{\frac{q^2-1}{q-1}}
-\left(C_2\frac{q^2-1}{q-1}+C_3\right)\frac{q^3-1}{q-1}  \ \ \Longrightarrow C_1=1,\  C_2=q^4,\ C_3=q^3
\label{exadeco}
\ee
Obviously exactly three parameters $C_j$ are needed in this case.
One of the morals of this example is that the first terms with small powers of one of the variables,
like those double-underlined in (\ref{exadeco0}),
are already the ``proper'' ones, but intermediate powers need deeper understanding of
internal structure of $\omega_N^{(a,m)}$, which could lead to decompositions like (\ref{exadeco}).
This is what we are going, to some extent, to reveal in the present paper.
We fully solved this problem for generic $\Omega_2^{(a,m)}$ at $N=2$, see (\ref{OmegaN2Q}),
making the trick (\ref{exadeco}) precise, and
for two series  at $N=3$: $\Omega_3^{(2,m)}$ with $a=2$, see (\ref{Omega32}),
and $\Omega_3^{(a,1)}$ with $m=1$, see (\ref{Ome3a1}).
Hopefully, further generalization of this curious exercise would seem interesting to someone else,
and more understanding would be achieved.

The $q\to 1$ limit of the twisted Cherednik system provides us with a simpler picture of the spectrum, however, it certainly does not allow one to make an unambiguous deformation. Indeed, in the untwisted case of $a=1$, the eigenfunctions of the Cherednik system are just non-symmetric Macdonald polynomials \cite{Opd95,Mac96,Che95,HHL,CO,BF97}, while the $q\to 1$ limit gives rise to the non-symmetric Jack polynomials \cite{BGHP,Opd95,KS}. It is not immediate to restore the former polynomials from the latter ones!

However, the insight obtained from limiting spectrum allows us to simplify calculations in the full twisted Cherednik system. In particular, it allows one to make a clever guess in order to generate the ground states $\Omega_N^{(a,m)}$.
Hence, our goal in this paper is to demonstrate how it works, first, in the case of the ground state, where we explain how to generate explicit formulas for $\Omega_N^{(a,m)}$, and, second, for the excitations. Note that the structure of these later has been already discussed in detail in \cite{MMPns}, however, here we give another look at them.

In section 2, we describe the twisted Cherednik system, its $q\to 1$ limit and describe the generic structures behind the limit and its deformation. In sections 3-4, we describe explicit calculations of the ground state $\Omega_N^{(a,m)}$, and, in section 5, discuss the first excitations. Section 6 contains some concluding remarks.

\subsection*{Notation}

At a given twist $a$, we use the auxiliary variables $X_i:=x_i^{1\over a}$ and $Q:=q^{1\over a}$, though
most parts of answers are expressed through $x_i$ and $q$. Indeed, intermediate formulas often
look much simpler in terms of $X_i$ and $Q$.

Throughout the paper, we use the quantum numbers
\be
[x]_q:={q^x-1\over q-1}
\ee
and the $q$-factorials
\be
[n]_q!:=\prod_{i=1}^n[i]_q
\ee
At $a=2$, we also sometimes use symmetric $Q$-quantum numbers:
\be
[n]_s:=\frac{Q^n-Q^{-n}}{Q-Q^{-1}}
\ee
This choice of quantum numbers can be convenient in some formulas to eliminate the negative powers of $Q$,
but this is reasonable only for the particular value of $a=2$.

We use upper-case letters for objects in the full twisted Cherednik system, and lower-case letters for their $q\to 1$ limit.

In section 5, we will also need the notation
\be\label{g}
g_n(x):=1+\frac{[n]_q(Q^a-1)}{1-x}
\ee

\section{General description}

\subsection{Twisted Cherednik system}

The $a$-twisted commuting Hamiltonians $\hat {\mathfrak{C}}_i^{(a)}$ of the twisted Cherednik \cite{Ch,NSCh} system are defined (see details in \cite{MMPns}\footnote{Notice that we shift the degree of $x_i$ in ${\cal C}_i^{(a)}$ as compared with \cite{MMPns} in order to have polynomial eigenfunctions. Otherwise, one has to multiply them by the factor of $q^{{1\over 2a}\sum_{i=1}^n(\log_q x_i)^2}$.})
\be
R_{ij}:&=&1+{(1-t^{-1})x_j\over x_i-x_j}(1-\sigma_{i,j})\\
R_{ij}^{-1}&=&1+{(1-t)x_j\over x_i-x_j}(1-\sigma_{i,j})\nn\\
{\cal C}_i^{(a)}:&=&
t^{1 - i} \left(\prod_{j = i + 1}^n R_{i,j}\right)
    {1\over x_i^{a-1\over a}}q^{\hat D_i}
    \left(\prod_{j = 1}^{i - 1} R^{-1}_{j,i}\right)\nn\\
   \hat {\mathfrak{C}}_i^{(a)}:&=&{1\over x_i}\Big(x_i{\cal C}_i^{(a)}\Big)^a
\ee
where $\hat D_i:=x_i{\p\over\p x_i}$. The products in ${\cal C}_i^{(a)}$ are arranged so that the smaller index stands to the left.

We put in these formulas $t=q^{-m}$, and are interested in the simultaneous eigenfunctions of these mutually commuting Hamiltonians:
\be
\hat {\mathfrak{C}}_i^{(a)}\cdot\Psi^{(a,m)}=\Lambda^{(a,m)}_i\cdot\Psi^{(a,m)},\ \ \ \ \ \ \ i=1,\ldots,N
\ee

\subsection{The $q\to 1$ limit}

The story begins at $q\longrightarrow 1$. Then, the commuting operators in the $q\to 1$ limit
($q = e^\hbar, \hbar \rightarrow 0$)
of the twisted Cherednik Hamiltonians are just
\be
\hat {\mathfrak{C}}_i^{(a)}\ \stackrel{q\to 1}{\longrightarrow}\
1 + \hbar
\hat {\mathfrak{ c}}^{(a,m)}_i
+ o(\hbar) = 1 + \hbar \cdot \omega_N^{(a,m)}\cdot {\cal D}_i\cdot \left(\omega_N^{(a,m)}\right)^{-1}
+ o(\hbar)
\ee
where
\be\label{Dl}
{\cal D}_i:=x_i{\p\over\p x_i}-m\sum_{i\ne j}{x_i\over x_i-x_j}(1-\sigma_{ij})-m\sum_{j>i}\sigma_{ij}
\ee
is the operator independent of $a$, $\sigma_{ij}$ is the operator of permutation of $x_i$ and $x_j$, and\footnote{Notice the connection with the Schur polynomials
$$
\omega_N^{(a,m)}
= {\rm Schur}_{[(N-1)(a-1),\ldots, 3(a-1),2(a-1),a-1]}^m[\vec X]
$$
which, at $a=2$, are exactly the $Q$-Schur polynomials studied in \cite{Suprun}
within the context of the Vogel's universality.}
\be
\omega_N^{(a,m)} = \left(\prod_{i<j}^N \frac{X_i^a-X_j^a}{X_i-X_j}\right)^m
\label{omega}
\ee
is just a function.  The eigenvalues similarly have perturbative expansion
$\Lambda_i^{(a,m)} = 1 + \hbar \lambda_{i}^{(a,m)} + o(\hbar)$. We are interested in the first order in $\hbar$.

The eigenfunctions of the operator ${\cal D}_i$ are the non-symmetric Jack polynomials \cite{BGHP,Opd95,KS} at $\beta=-m$, $J_\alpha^{(m)}$, and the common eigenfunctions of operators $\hat {\mathfrak{ c}}^{(a,m)}_i$,
\be
\hat {\mathfrak{ c}}^{(a,m)}_i  \psi^{(a,m)}_\alpha = \lambda^{(a,m)}_{\alpha,i} \psi^{(a,m)}_\alpha
\label{clatwChe}
\ee
are labeled by \wcs\ $\alpha$ with $N$ parts
and have the spectacularly simple form:
\be
\psi^{(a,m)}_\alpha = \omega_N^{(a,m)}[\vec X]\cdot J_\alpha^{(m)}[\vec x]
\label{zdeco}
\ee

The non-symmetric Jack polynomials can be certainly obtained from the non-symmetric Macdonald polynomials in the limit of $q\longrightarrow 1$.

The two ``extremal'' \wcs\ in the set of possible $\alpha$'s are
\be
J^{( m)}_{[L,0,\ldots,0]} \sim  \sum_{k,j_2,\ldots,j_N=0}^L  \delta_{k+j_2+\ldots+j_N,L}\cdot   \frac{ x_1^{k}}{k!(m-1-k)!}\prod_{s=2}^N \frac{x_s^{j_s}}{j_s! (m-j_s)!}
\label{eL0}
\ee
and
\be
J^{( m)}_{[0,\ldots,0,L]} \sim   \sum_{k,j_2,\ldots,j_N=0}^{L-1}  \delta_{k+j_2+\ldots+j_N,L-1}\cdot   \frac{ x_1^{k}}{k!(m-k)!}\left(\prod_{s=2}^{N-1} \frac{x_s^{j_s}}{j_s! (m-j_s)!}\right)   \frac{x_N^{j_N+1}}{j_N! (m-j_N-1)!}
\label{e0L}
\ee
The asymmetry of the first polynomial touches only $x_1$, in other variables it is symmetric.
In the second polynomial, an additional asymmetry concerns $x_N$.

\bigskip

\begin{figure}[h]

\begin{picture}(100,220)(-230,0)

\put(-200,0){\line(1,0){400}}
\put(0,0){\line(0,1){200}}
\put(0,198){\vector(0,1){2}}

\put(-150,100){\line(1,0){200}}
\put(-150,150){\line(1,0){200}}
\put(-30,151){\line(1,0){60}}
\put(-28,151){\vector(-1,0){2}}
\put(28,151){\vector(1,0){2}}

\qbezier(0,100)(10,170)(150,200)
\qbezier(0,100)(-10,170)(-150,200)

\put(10,155){\mbox{$n_L$}}
\put(-90,120){\mbox{$L$}}
\put(-185,50){\mbox{$L_0=\frac{N(N-1)}{2}\cdot(\alpha-1)m$}}

\put(-70,0){\line(0,1){150}}

\put(-70,2){\vector(0,-1){2}}
\put(-70,98){\vector(0,1){2}}
\put(-70,102){\vector(0,-1){2}}
\put(-70,148){\vector(0,1){2}}

\put(70,145){\mbox{$\sum_L n_Lq^L = \frac{1}{(1-q)^{N}}$}}
\put(-65,200){\mbox{\text{ degree in $\vec X$}}}

\end{picture}

\caption{\footnotesize The spectrum of common polynomial eigenfunctions.
Plotted on the vertical axis are the degrees of polynomials in $X_i$.
There is a single ground state eigenfunction $\omega_N^{(a,m)}$ of degree $\frac{N(N-1)}{2}(\alpha-1)m$.
Over it, there are the ``excitations'' of additional degrees $L$.
They are labeled by \wcs, so that there are $n_L$ different polynomials of the additional degree $L$.
For particular values of $m$, there can be degeneracies, and some of these polynomials coincide, as well as their eigenvalue sets.
Eigenvalues are not shown.}

\label{spectrum}
\end{figure}
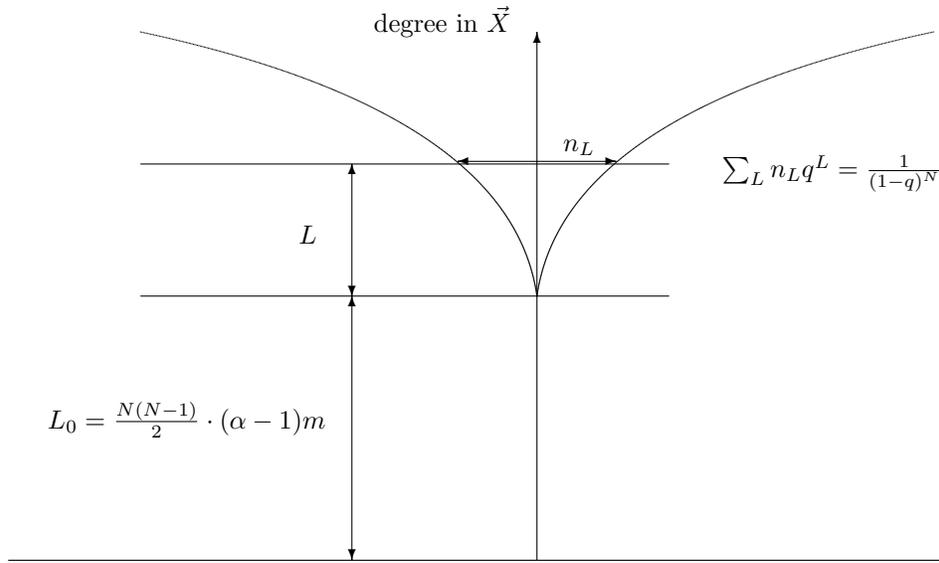

Thus, the spectrum of polynomial eigenfunctions is somewhat peculiar, see Fig.\ref{spectrum}:
there is a gap $L_0:=\frac{N(N-1)}{2}\cdot(\alpha-1)m$ in the grading of the ground state, and, above the ground state polynomial, there are as many excitations of grading $L_0+L$ as there are different non-symmetric graded polynomials of
degree $L$.

To understand the origins of this pattern, one can look at the simplest case of $N=2$.
Then the two commuting operators are:
\be
\hat {\mathfrak{c}}_1^{(a,m)} = X_1 \frac{\p}{\p X_1} + \frac{mX_2}{X_1-X_2}\hat \sigma_{12} + \frac{X_2^a}{X_1^a-X_2^a}  + {\rm const},
\nn \\
\hat {\mathfrak{c}}_2^{(a,m)} = X_2 \frac{\p}{\p X_2} - \frac{mX_2}{X_1-X_2}\hat \sigma_{12} + \frac{X_1^a}{X_1^a-X_2^a}  + {\rm const}
\label{quaclaChe12}
\ee
Clearly these operators acting on unity (a polynomial of degree 0) produce a polynomial again only at $a=1$.
At $a>2$, the only chance to get a polynomial by cancellation of two terms with the denominators
$X_1-X_2$ and $X_1^a-X_2^a$ is to act on a polynomial of degree at least $a-1$.
But more detailed analysis raises the degree further to $(a-1)m$.

For $N=2$, formulas (\ref{eL0}) and (\ref{e0L})   actually provide a complete description of excitations,
because all other diagrams are of the form either $\alpha = [L'+l,l]$ or $\alpha = [l,l+L']$ with $L=L'+2l$,
and the corresponding non-trivial Jack parts of eigenfunctions are just
\be
J^{(m)}_{[L'+l,l]} = \left(\prod_{i=1}^2x_i\right)^l\cdot J^{(m)}_{[L',0]}, \nn \\
J^{(m)}_{[l,L'+l]} = \left(\prod_{i=1}^2x_i\right)^l\cdot J^{(m)}_{[0,L']}
\label{ereduct}
\ee
i.e. trivially depend on $l$ (eigenvalues, however, depend on both $\alpha$ and $l$).

However, generic excitations for $N>2$ are much trickier, though the same property (\ref{ereduct}) preserves:
\be
J^{(m)}_{[\alpha_1+l,\alpha_2+l,\ldots,\alpha_N+l]} = \left(\prod_{i=1}^nx_i\right)^l\cdot J^{(m)}_{[\alpha_1,\alpha_2,\ldots,\alpha_N]}
\label{ereduct2}
\ee
Formulas (\ref{ereduct}), (\ref{ereduct2}) (and (\ref{Zreduct}) below) follow from the fact that
\be\label{fac}
{\cal C}_i\Big(\prod_{i=1}^Nx_i^\alpha\cdot F(x)\Big)=q^\alpha \prod_{i=1}^Nx_i^\alpha\cdot{\cal C}_iF(x)
\ee
and, hence, multiplying an eigenfunction of ${\cal C}_i$, $\hat {\mathfrak{C}}_i^{(a)}$, or ${\cal D}_i$ by $\prod_{i=1}^Nx_i^\alpha$ maps the eigenfunction to an eigenfunction.

\subsection{On the principles and the {\it base} of $q$-deformation}

In the case of quantum groups, deformation of many expression is basically the change of factorials for $q$-factorials
and insertion of powers of $q$, which at most quadratic in the relevant variables (like $a$, $m$, $L$ or $j$).
We call {\it such} $q$-deformations {\it direct}.
However, this is true only for a specially cooked $q\to 1$ expression, which we call a {\it base}
for deformation.
It usually requires
appropriate decompositions, involving splitting naive sums into pieces,
i.e. increasing the number of summations.
No {\it a priori} recipe is known for such a splitting, but it is convenient to {\it formulate} the answer
at least {\it a posteriori} by providing such splitting.

In our case, there will be two splittings of this kind.
First, the product (\ref{zdeco}) will be substituted by a sum:
\be
\Psi_\alpha^{(a,m)} = \sum_{\sigma}  \Xi_{\sigma}^{(a,m)}[\vec X]\cdot E_\alpha^{\sigma}[\vec x]
\label{Zdeco0}
\ee
where $\Xi_{\sigma}^{(a,m)}$ is proportional to $\Omega_N^{(a,m)}$ at some shifted point in $X_i$'s and maybe at some shifted value of $m$, multiplied by a monomial of $X_i$'s, while $E_\alpha^{\sigma}[\vec x]$ are some coefficient functions of $\vec x$ (not of $\vec X$) independent of $a$.
Second, $\Omega^{(a,m)}$ itself is a $Q$-deformation of a skillfully decomposed product (\ref{omega}).

Both decompositions are currently completely worked out for $N=2$, generalization to other $N$ is also known though in a less explicit form, see \cite{MMPns} for details.

\bigskip

However, even for $N=2$ the base for {\it direct} $Q$-deformation is non-trivial.
Instead of (\ref{eL0}) and (\ref{e0L}) we should use another decomposition
(even its equality to (\ref{eL0}) and (\ref{e0L}) is not obvious, see \cite[Appendix]{MMPns} for details):
\be
J_{[L,0]}^{(m)} = \sum_{j=0}^L
 \frac{L!}{j!(L-j)!}\frac{(m-L-1)!}{(m-j-1)!}\frac{(2m-j-1)!}{(2m-L-1)!} \,(x_1-x_2)^j x_2^{L-j}
 \label{e2L0alt}
\ee
and
\be
J_{[0,L]}^{(m)} =  \sum_{j=0}^{L-1}
 \frac{(L-1)!}{j!(L-j-1)!}\frac{(m-L)!}{(m-j-1)!}\frac{(2m-j-1)!}{(2m-L)!} \,(x_1-x_2)^j x_2^{L-j}
\label{e20Lalt}
\ee

Likewise, $\omega$ should be converted to the form
\be
\left(\frac{X_1^a-X_2^a}{X_1-X_2}\right)^m
= \sum_{j=0}^{(a-1)m} X_1^jX_2^{(a-1)m-j} \sum_{k=0}^{{\rm floor}\left(\frac{j}{a}\right)}
(-)^k \underbrace{\frac{m!}{k!(m-k)!} \frac{(m+j-ak-1)!}{(m-1)!(j-ak)!}}_{
m\cdot  \frac{(m+j-ak-1)!}{k!(m-k)!(j-ak)!}
}
\label{amundeformed}
\ee
Note that even for $a=2$ this is a non-conventional representation of $(X_1+X_2)^m$.

\paragraph{Expansion (\ref{amundeformed}) gets clear from the following example.}

\be
\omega_2^{(4,3)} &=& (X_1^3+X_1^2X_2+X_1X_2^2+X_2^3)^3  = X_1^9 + 3X_1^8X_2+6X_1^7X_2^5+10X_1^6X_2^3+(15-3)X_1^5X_2^4+\nn\\
&+&(21-9)X_1^4X_2^5+(28-18)X_1^3X_2^6+(36-30)X_1^2X_2^7+(45-45+3)X_1X_2^8
+(55-63+9)X_2^9\Longrightarrow\nn\\
&\Longrightarrow& \sum_{i=0}^9 \left({(i+1)(i+2)\over 2}-3\ {(i-2)(i-3)\over 2}-3\ {(i-6)(i-7)\over 2}\right)X_1^iX_2^{9-i}\Longrightarrow\nn\\
&\Longrightarrow& \sum_{i=0}^{3m}\left(\frac{(m+i-1)!}{(m-1)!i!}-m \frac{(m+i-5)!}{(m-1)!(i-4)!}\frac{m(m-1)}{2} \frac{(m+i-9)!}{(m-1)!(i-8)!}
\right)X_1^iX_2^{3m-i}
\Longrightarrow\nn\\
&\Longrightarrow&
\sum_{i=0}^{(a-1)m} X_1^iX_2^{(a-1)m-i} \sum_{k=0} (-)^k \frac{m!}{(m-k)!k!} \frac{(m+i-1-ak)!}{(m-1)!(i-ak)!}
\ee

\bigskip

The same logic applies to products of several $\omega$, e.g.
\be
\omega_2^{(a_1,m_1)}\omega_2^{(a_2,m_2)} =
\sum_{i=0}^{(a_1-1)m_1+(a_2-1)m_2} X_1^iX_2^{(a_1-1)m_1+(a_2-1)m_2-i}
\cdot \nn \\ \cdot
\sum_{k_1,k_2=0} (-)^{k_1+k_2}  \frac{m_1!}{(m_1-k_1)!k_1!} \frac{m_2!}{(m_2-k_2)!k_2!}
\frac{(m_1+m_2+i-1-a_1k_1-a_2k_2)!}{(m_1+m_2-1)!(i-a_1k_1-a_2k_2)!}
\label{omeome}
\ee

\noindent
and, generally,
\be
\prod_{s=1}^S \omega_2^{(a_s,m_s)} =
\sum_{i=0}^{A} X_1^iX_2^{A-i}
\sum_{\vec k}   \left( \prod_{s} (-)^{k_s} \frac{m_s!}{(m_s-k_s)!k_s!}\right)
\frac{\Big(i-1+\sum_s (m_s-a_sk_s) \Big)!}{\Big(-1+\sum_s m_s\Big)!\Big(i-\sum_s a_sk_s\Big)!}
\label{omeS}
\ee
where $A = \sum_s (a_s-1)m_s$ and the sum over $k_s$ is cut off by factorials in the denominators: hereafter, we understand
\be
{1\over k!}:={1\over \Gamma(k+1)}\Longrightarrow 0\ \ \ \ \ \hbox{at}\ -k\in\mathbb{N}
\ee

In the next two sections, we apply this knowledge to $N>2$ in two rather different cases:
to the case of minimal non-trivial twist $a=2$,
and to an arbitrary twist $a>2$.
The difference is that, in the first case, $\omega_N^{(2,*)}$
is always a {\it linear} combination of $\omega_2^{(2,*)}$,
thus the main part of deformation is the change for $\Omega_2^{(2,*)}$,
which we already know from (\ref{OmegaN2Q}).
In the second case, one needs a $Q$-deformation of {\it multi}-linear combinations
of $\omega_2^{(a,*)}$,
which is, to some extent, implied by (\ref{omeome}) and (\ref{omeS}),
still needs to be worked out.
We do this by requiring that $Q$-deformation of (\ref{omeome}) solves (\ref{peri})
and then by adjusting the remaining $X_3,\ldots,X_N$-dependent coefficients
from the symmetry property.
In the both cases of $a=2$ and $a>2$, we did not work out fully generic formulas,
but two series are already available at $N=3$:
$\Omega_3^{(2,m)}$ and $\Omega_3^{(a,1)}$, as well as examples of $\Omega_3^{(3,m)}$.

However, before moving to consideration of $N>2$, we first briefly describe the $N=2$ case.

\subsection{A simple example: deformation at $N=2$}

We describe the common eigenfunctions of the twisted Cherednik operators,
\be
\hat {\mathfrak{C}}^{(a)}_i  \Psi^{(a,m)}_\alpha = \Lambda^{(a,m)}_{\alpha,i} \Psi^{(a,m)}_\alpha,\ \ \ \ \ \ \ i=1,2
\ee
Eigenvalues are immediately inherited from those in the $q\to 1$ limit,
\be
\Lambda^{(a,m)}_{\alpha,i} = Q^{\lambda^{(a,m)}_{\alpha,i}}
\ee
and the reduction property (\ref{ereduct}) is still true:
\be
\Psi^{(a,m)}_{[L'+l,l]} = \left(\prod_{i=1}^2x_i\right)^l\Psi^{(a,m)}_{[L',0]}, \nn \\
\Psi^{(a,m)}_{[l,L'+l]} = \left(\prod_{i=1}^2x_i\right)^l\Psi^{(a,m)}_{[0,L']}
\label{Zreduct}
\ee
At $N=2$, the eigenfunctions (\ref{Zdeco0}) for the two \wcs\ $\alpha=[L,0]$ and $[0,L]$ take the form
\be
\!\!\!\!\!\!\!\!\!\!\!\!\!\!\!\!\!\!\!\!\Psi^{(a,m)}_{[L,0]} = \sum_{j=0}^L  C\cdot
 \frac{[L]_q!}{[j]_q![L-j]_q!}\frac{[m-L-1]_q!}{[m-j-1]_q!}\frac{[2m-j-1]_q!}{[2m-L-1]_q!}
\cdot \Omega^{(a,m-j)}_2\Big(X_1,X_2Q^{L-j}\Big)X_2^{L-j}\prod_{i=0}^{j-1} (q^{i-m}x_1-x_2)
\label{ZL0}
\ee
and
\be
\!\!\!\!\!\!\!\!\!\!\!\!\!\!\!\!\!\!\!\!\Psi^{(a,m)}_{[0,L]} =   \sum_{j=0}^{L-1} C \cdot
\frac{[L-1]_q!}{[j]_q![L-j-1]_q!}
\frac{[m]_q}{[m-j]_q} \frac{[m-L]_q!}{[m-j-1]_q!}\frac{[2m-j-1]_q!}{[2m-L]_q!}
\cdot \Omega^{(a,m-j)}_2\Big(X_1,X_2Q^{L-j}\Big)X_2^{L-j}
\prod_{i=0}^{j-1} (q^{i-m}x_1-x_2)\nn\\
\label{Z0L}
\ee
with $C=Q^{j^2-\frac{aj(j-1)}{2}+(2a-1)mj-Lj}$.
These $\Psi^{(a,m)}_\alpha$ are polynomials of degree $(a-1)(m-j)+L-j+aj = (a-1)m+L$ in $X$.
In an obvious sense, they are {\it direct} $Q$-deformations of
(\ref{e2L0alt}) and (\ref{e20Lalt}).
Note that all quantum numbers are asymmetric and depend on $q$, not on $Q$, but additional $Q$-powers are not always
integer powers of $q=Q^a$.

According to these formulas, all $a$-dependence is concentrated in a single function $\Omega$
and in monomial $Q$- and $X$-factors, this is what we denoted $\Xi_{\sigma}^{(a,m)}$ in (\ref{Zdeco0}).
As to the function $\Omega$, it is a {\it direct} $Q$-deformation of (\ref{amundeformed})
with the only subtlety: a non-trivially decomposed factor $m$:
\be\label{34}
\Omega^{(a,m)}_2
 &=& \sum_{j=0}^{(a-1)m} Q^{j^2-(a-1)mj}  X_1^jX_2^{(a-1)m-j}
\times \nn \\ &\times&
\sum_{k=0}^{\frac{j}{a}}
q^{(a-1)mk-(j-1)k+\frac{k(k-1)}{2}}
\frac{[m+j-ak-1]_q!}{  [j-ak]_q![m-k]_q![k]_q!}
 \left([m-k]_q + q^{j+m-(a+1)k} [k]_q\right)
\label{OmegaN2Q}
\ee
Note that any dependence on the fractional power of $q$, i.e. on $Q$ remains only in the
weight factor at the very beginning of the formula.

\section{The ground state $\Omega^{(2,m)}_N$
\label{sec_a=2}}

\subsection{$N=2$}

In constructing $\Omega^{(2,m)}_N$, we are going to obtain and use recursion formulas, and start with the simplest case of $N=2$ when we know the answer (\ref{OmegaN2Q}). The recursion formula in this case is
\be\label{eq:cut-n-join-like-rec}
\Omega_2^{(2,m+2)} = \Omega_2^{(2,m)}\Omega_2^{(2,2)} + [m+2]_Q[m]_Q\Omega_2^{(2,m)} X_1X_2 \frac{(Q-1)^2}{Q^{m+1}}
\ee
which , in its structure is reminiscent of cut-and-join equations
\textit{a la} \cite{paper:MMN-integrability-hurwitz-II}.

\subsection{$N=3$}

In this case, the $Q=1$ formula looks like
\be
\omega_3^{(2,1)} =\underbrace{(X_1+X_2)}_{\omega_2^{(2,1)}}(X_1+X_3)(X_2+X_3) = \Big((X_3^2+X_1X_2)+X_3\omega^{(2,1)}_2\Big)\omega^{(2,1)}_2
\ee
and implies that
\be
\omega_3^{(2,m)} =(X_1+X_2)^m(X_1+X_3)^m(X_2+X_3)^m
= \Big((X_3^2+X_1X_2)+X_3\omega^{(2,1)}_2\Big)^m\omega^{(2,m)}_2
= \nn\\
= \sum_{j=0}^m \frac{m!}{(m-j)!j!} X_3^j(X_3^2+X_1X_2)^{m-j}\omega^{(2,m+j)}_2
\label{omega32}
\ee
Here and below, $\omega^{(a,m)}_2$ and $\Omega^{(a,m)}_2$ denote functions of the first two variables, $X_1$ and $X_2$ if opposite is not indicated.

In order to understand how to deform this formula, consider a few examples:
\be
\Omega_3^{(2,1)} = (X_3^2+X_1X_2)\Omega_2^{(2,1)} + X_3\Omega_2^{(2,2)}
= {\cal X}_0\Omega_2^{(2,1)}+X_3\Omega_2^{(2,2)}
\nn \\ \nn \\
\Omega_3^{(2,2)} = (X_3^2+X_1X_2)^2\Omega_2^{(2,2)} + (Q+Q^{-1})(X_3^2+X_1X_2)X_3\Omega_2^{(2,3)}
+ X_3^2\left(\Omega_2^{(2,4)} + \frac{(Q-1)^2}{Q}X_1X_2\Omega_2^{(2,2)}\right)
= \nn\\ =
{\cal X}_1 \Omega_2^{(2,2)} + [2]_sX_3{\cal X}_0\Omega_2^{(2,3)} + X_3^2\Omega_2^{(2,4)}
\nn \\ \nn \\
\Omega_3^{(2,3)} = (X_3^2+X_1X_2)^3\Omega_2^{(2,3)} +(Q^2+1+Q^{-2})(X_3^2+X_1X_2)^2 X_3\Omega_2^{(2,4)}
+ \nn \\
+ (X_3^2+X_1X_2) X_3^2\left((Q^2+1+Q^{-2})\Omega_2^{(2,5)} + \frac{(Q^2-1)^2}{Q^2}X_1X_2\Omega_2^{(2,3)}\right)
+ \nn \\
+ X_3^3\left(\Omega_2^{(2,6)}+(Q^2+1+Q^{-2})\frac{(Q-1)^2}{Q}X_1X_2\Omega_2^{(2,4)}\right)
= \nn \\ =
{\cal X}_0{\cal X}_2 \Omega_2^{(2,3)} + [3]_sX_3{\cal X}_1\Omega_2^{(2,4)} +[3]_sX_3^2{\cal X}_0\Omega_2^{(2,5)}
+ X_3^3\Omega_2^{(2,6)}\nn
\ee
\be
\Omega_3^{(2,4)} = (X_3^2+X_1X_2)^4\Omega_2^{(2,4)}
+(Q^2+Q^{-2})(Q+Q^{-1})(X_3^2+X_1X_2)^3 X_3\Omega_2^{(2,5)}
+ \nn \\
+ (X_3^2+X_1X_2)^2 X_3^2\left((Q^2+Q^{-2})(Q^2+1+Q^{-2})\Omega_2^{(2,6)} + \left(\frac{(Q^3-1)^2}{Q^3}+\frac{(Q-1)^2}{Q}\right)X_1X_2\Omega_2^{(2,4)}\right)
+ \nn \\
+ (X_3^2+X_1X_2) X_3^3 (Q^2+Q^{-2})(Q +Q^{-1})\left(\Omega_2^{(2,7)} + \frac{(Q^2-1)^2}{Q^2}X_1X_2\Omega_2^{(2,5)}\right)
+ \nn \\
+ X_3^4\left(\Omega_2^{(2,8)}+(Q^2+Q^{-2})(Q^2+1+Q^{-2})\frac{(Q-1)^2}{Q}X_1X_2\Omega_2^{(2,6)}
+\frac{(Q^3-1)^2}{Q^3}\frac{(Q-1)^2}{Q}(X_1X_2)^2\Omega_2^{(2,4)}\right)
= \nn \\ =
{\cal X}_1{\cal X}_3\Omega_2^{(2,4)} + [4]_sX_3{\cal X}_0{\cal X}_2\Omega_2^{(2,5)}+
\frac{4]_s[3]_s}{2}X_3^2{\cal X}_1\Omega_2^{(2,6)} + [4]_sX_3^3{\cal X}_0\Omega_2^{(2,7)} + X_3^4\Omega_2^{(2,8)}
\nn\\
\Omega_3^{(2,5)} = {\cal X}_0{\cal X}_2{\cal X}_4\Omega_2^{(2,5)}
+  [5]_s{\cal X}_1{\cal X}_3 X_3\Omega_2^{(2,6)}
+  \frac{[5]_s[4]_s}{[2]_s} {\cal X}_0{\cal X}_2X_3^2\Omega_2^{(2,7)}
+  \frac{[5]_s[4]_s}{[2]_s}  {\cal X}_1X_3^3\Omega_2^{(2,8)}
+  [5]_s {\cal X}_0  X_3^4\Omega_2^{(2,9)} + X_3^5\Omega_2^{(2,10)}
\nn
\ee
Here
\be
{\cal X}_s := \Big(X_3^2+X_1X_2\Big)^2 +  \frac{(Q^s-1)^2}{Q^s}X_1X_2X_3^2
= X_3^4+(Q^s+Q^{-s})X_3^2X_1X_2+(X_1X_2)^2
\label{callX}
\ee
According to this definition, ${\cal X} = \sqrt{{\cal X}_0}$.

\bigskip

In general
\be
\boxed{
\Omega_3^{(2,m)} = \sum_{j=0}^m  \frac{[m]_s!}{[m-j]_s![j]_s!} {\cal Y}_{m-j}  X_3^j\Omega_2^{2,m+j}
}
\label{Omega32}
\ee
Specified in this formula should be only the ${\cal Y}$-factors.
Note that they do not depend on $m$ and $j$ separately, only on $m-j$, and $Y_{m-j} \sim {\cal X}^{m-j}$.
From above examples we read:
\be
{\cal Y}_m = \prod_{i=1}^m (X_3^2+Q^{m+1-2i}X_1X_2)
\ee
Together with (\ref{callX}) and (\ref{OmegaN2Q})
this promotes (\ref{Omega32}) to a full formula for $\Omega$ at $N=3$, $a=2$ and arbitrary $m$.
Once found, it can be easily checked to be, in fact,
an eigenfunction for quite large values of $m$.
This looks like {\bf a typical NP problem}: the equations are difficult to solve, but easy to check.

Since (\ref{Omega32}) is linear in $\Omega_2$ multiplied by a polynomial of the product $X_1X_2$, it automatically satisfies
the periodicity equations (\ref{peri}) at $k=1$, $l=2$.
Instead, a non-trivial property of (\ref{Omega32}) is that this expression is a symmetric polynomial of $X_1$, $X_2$ and $X_3$.
We emphasize that this {\it symmetry} is not obvious neither in expression (\ref{Omega32})
nor in its predecessor (\ref{omega32}) (where, however, it follows from its symmetric origin).

\subsection{$N>3$}

The same trick works at $a=2$ at arbitrary $N$.
In this case, $\Omega_N^{(2,m)}$ can always be expanded in a linear combination of $\Omega_2$ with coefficients polynomial in the product $X_1X_2$,
and thus satisfies periodicity equations in the first two variables.
After that, the coefficients can be restored from the symmetry requirement, and the answer is
automatically an eigenfunction.
In order to demonstrate how this works, we provide just a couple of additional examples, at $N=4$ and $m=1,2$ ($m=1$ belongs also to another series):
\be\label{41}
\Omega_4^{(2,1)} = (X_3+X_4)(X_3^2+X_1X_2)(X_4^2+X_1X_2)\,\Omega_2^{(2,1)}
+ \nn \\
+ \Big(X_3+QX_4\Big)\Big(X_3+ \frac{1}{Q}X_4\Big)(X_3X_4+X_1X_2)\,\Omega_2^{(2,2)}
+ (X_3+X_4)X_3X_4\,\Omega_2^{(2,3)} = \nn \\
= X_1X_2\,\Omega_{X_1,X_2}^{(2,1)}\Omega_{X_3,X_4}^{(2,3)}
+(X_1X_2+X_3X_4)\,\Omega_{X_1,X_2}^{(2,2)}\Omega_{X_3,X_4}^{(2,2)}
+ X_3X_4\,\Omega_{X_1,X_2}^{(2,3)}\Omega_{X_3,X_4}^{(2,1)} - \nn \\
- \Big(X_1X_2-Q^2X_3X_4\Big)\Big(X_1X_2-\frac{1}{Q^2}X_3X_4\Big)\,\Omega_{X_1,X_2}^{(2,1)}\Omega_{X_3,X_4}^{(2,1)}
\ee
Explicit in the last version is symmetry between the pairs $X_1,X_2$ and $X_3,X_4$ and within them.
Symmetry with respect to all the permutations of $X_1,X_2,X_3,X_4$ is, however, not explicit, but true.

Similarly,
{\footnotesize
\be
\Omega_{N=4}^{(2,2)} =
\left((X_1^2X_2^2+X_3^2X_4^2)^2 + [2]_sX_1X_2(X_1^2X_2^2+X_3^2X_4^2)(X_3^2+X_4^2)
+X_1^2X_2^2\Big(X_3^2+Q^2X_4^2\Big)\Big(X_3^2+\frac{1}{Q^2}X_4^2\Big)\right)
\,\Omega_{X_1,X_2}^{((2,2)}\Omega_{X_3,X_4}^{((2,2)}
+ \nn \\
+ [2]_s\Big(X_1^2+X_2^2+X_3^2X_4^2+X_1X_2(X_3^2+X_4^2)\Big)\,\Omega_{X_1,X_2}^{((2,3)}\Omega_{X_3,X_4}^{((2,3)}
+ \nn \\
\!\!\!\!\!\!\!\!
+ \Big(X_3^2X_4^2(X_3^2+X_4^2)+[4]_sX_1X_2X_3^2X_4^2+X_1^2X_2^2(X_3^2+X_4^2)
+ [4]_sX_3^3X_4^3+[2]_s^2X_1X_2X_3X_4(X_3^2+X_4^2)+[4]_sX_1^2X_2^2X_3X_4\Big)
\,\Omega_{X_1,X_2}^{((2,4)}\Omega_{X_3,X_4}^{((2,2)}
+ \nn \\
+ [2]_sX_3X_4(X_1X_2+X_3X_4)\,\Omega_{X_1,X_2}^{((2,5)}\Omega_{X_3,X_4}^{((2,3)}
+X_3^2X_4^2\,\Omega_{X_1,X_2}^{((2,6)}\Omega_{X_3,X_4}^{((2,2)}
\nn
\ee
}

Extension to higher $N$ requires technical tricks in order to simplify calculations.
The simplest option is to look at the expansion in the form
\be
\Omega_N^{(2,m)} = \sum_{i,j,R}  u_{ij}^R\cdot (X_1X_2)^i \Omega_{X_1,X_2}^{(2,j)} S_R[\vec X]
\ee
where $S_R[\vec X]$ denotes the symmetric Schur polynomials of variables $X_3,\ldots,X_N$.
The size of the Young diagrams $R$ is restricted by the condition $i+j+|R|=\frac{mN(N-1)}{2}$,
and $j$ actually runs from $m$ to $m(N-1)$.
Restriction to  $a=2$ allows one to take this anzatz linear in $\Omega_2$, thus it automatically
satisfies {\it periodicity} equations w.r.t. $X_1$ and $X_2$.
Coefficients $u$ are restored from the symmetry between, say, $X_1$ and $X_3$, though
this means solving a rather large system of linear equations.
As an example, one obtains for the simplest $\Omega$ at $N=5$
\be
\Omega_{N=5}^{(2,1)} =
\left\{ (S_{[2,3,4]}+X_1X_2S_{[3,4]}+X_1^2X_2^2S_{[1,4]}+X_1^3X_2^3S_{[1,2]})
 +     \phantom{\frac{(Q-1)^2}{Q}}    \right. \nn \\ \left.
+ \frac{(Q-1)^2}{Q}\Big(S_{[3,3,3]}+X_1X_2(S_{[1,3,3]}-S_{[2,2,3]})
+X_1^2X_2^2(S_{[1,1,3]}-S_{[1,2,2]})+X_1^3X_2^3S_{[1,1,1]}\Big)
\right\}
\Omega_{X_1,X_2}^{(2,1)}
+\nn\\ \nn \\
 +\left\{ \Big(S_{[1,3,4]}+X_1X_2S_{[2,4]}+X_1^2X_2^2S_{[1,3]}\Big)
 + \frac{Q^2-Q+1}{Q}\Big(S_{[2,2,4]}+X_1X_2(S_{[3,3]}+S_{[1,1,4]}) + X_1^2X_2^2S_{[2,2]}\Big)
 +\right. \nn \\ \left.
 + \frac{(Q^2-Q+1)^2}{Q^2}X_1X_2S_{[2,2,2]}
 + \frac{Q^4-Q^3+Q^2-Q+1}{Q^2}(S_{[2,3,3]}+X_1^2X_2^2S_{[1,1,2]})
\right\}
\Omega_{X_1,X_2}^{(2,2)}
+\nn
\ee
\be
 +\left\{
(S_{[1,2,4]}+X_1X_2S_{[2,3]})+\frac{Q^2-Q+1}{Q}(S_{[1,3,3]}+X_1X_2S_{[1,1,3]})
+ \frac{Q^4-Q^3+Q^2-Q+1}{Q^2}(S_{[2,2,3]}+X_1X_2S_{[1,2,2]}
\right\}
\Omega_{X_1,X_2}^{(2,3)}
+\nn
\ee
\be\label{43}
+ \left\{S_{[1,2,3]} + \frac{(Q-1)^2}{Q^2} S_{[2,2,2]}\right\}\Omega_{X_1,X_2}^{(2,4)}
\ee

\section{The ground state $\Omega^{(a,m)}_N$}

\subsection{$N=3$, $m=1$ at any $a$}

The simplest next case is equally simple:
\be
\Omega_3^{(3,1)} = B_2\Omega_2^{(3,1)}
+\eta X_3B_1\Omega_2^{(2,1)}\left( \frac{1}{Q}\Omega_2^{(3,1)}+X_1X_2\frac{(Q^2-1)(Q-1)}{Q^3}\right)
+ X_3^2\Omega_2^{(3,2)}, \ \ \ \ \ \ \boxed{\eta=1}
\label{Omega331}
\ee
where we introduced the quantity
\be
B_k:=\left\{\begin{array}{ccc}  X_3^{2k}+(X_1X_2)^k && k>0 \\ 1 && k=0 \end{array} \right.
\ee
This expression satisfies the periodicity equations in $X_1$, $X_2$ at arbitrary $\eta$, but is {\it symmetric} and becomes an eigenfunction at a particular choice of the coefficient $\eta=1$.

However, for higher $a$ and $m$, this fails to work in such a simple way.
The periodicity equations  in $X_1$, $X_2$ are still much simpler to solve, however, finding a proper anzatz that admits a symmetric polynomial is not that simple. The right anzatz for an arbitrary $a$ is actually similar to (\ref{OmegaN2Q}).
It is implied by (\ref{omeome}), which in this case is $Q$-deformed {\it directly} , and gives
\be
\!\!\!\!\!\!\!\!\!
\boxed{
\begin{array}{c}
\Omega_{N=3}^{a,1} = B_{a-1}\Omega_2^{(a,1)} + X_3^{a-1}\Omega_2^{(2,2)}
+  \\ \\
+  \sum_{b=2}^{a-1} X_3^{b-1}B_{a-b}\left\{
\sum_{c=0}^{b-2}\eta_{a,b,c} [c+1]_q \Big(X_1^cX_2^{a+b-2-c}+X_1^{a+b-2-c}X_2^c\Big)
+ \sum_{c=b-1}^{a-1} \eta_{a,b,c} [b]_q X_q^cX_2^{a+b-2-c}
\right\}
\end{array}
}
\label{Ome3a1}
\ee
with
\be
\eta_{a,b,c} = Q^{c^2-(b+c-1)a-(b-2)c+b(b-1)}
\ee

\subsection{  $m=1,2$ at $a=3$}

Now we fix $a$ at the first really non-trivial value, $a=3$ and increase $m$.

The case of ${\bf m=1}$ is straightforward:
\be
\omega_3^{(3,1)} = B_2\,\omega_2^{(3,1)}
&+ X_3B_1\,\omega_2^{(2,1)}\omega_2^{(3,1)}
&+ X_3^2\,\omega_2^{(3,2)}
\nn \\ \nn \\
&\downarrow
\nn \\ \nn \\
\omega_3^{(3,1)} = (X_3^4+X_1^2X_2^2)\,\omega_2^{(3,1)}
&+ X_3(X_3^2+X_1X_2)\,\omega_2^{(2,1|3,1)}
&+ X_3^2\,\omega_2^{(3,2)}
\nn \\ \nn \\
&{\downarrow}
\nn
\ee
\be
\boxed{
\Omega_3^{(3,1)} = (X_3^4+X_1^2X_2^2)\cdot\Omega_2^{(3,1)}
+ \frac{1}{Q}X_3(X_3^2+X_1X_2)\cdot\Omega_2^{(2,1|3,1)}
+ X_3^2\cdot\Omega_2^{(3,2)}
}
\label{Omega331p}
\ee
where we introduced the evident notation
\be
\omega_2^{(2,1|3,1)} := \omega_2^{(2,1 )} \omega_2^{( 3,1)}
&= (X_1^3+X_2^3)+2X_1X_2(X_1+X_2)
\nn \\
\Omega_2^{(2,1|3,1)}:=\Omega_2^{(2,1 )} \Omega_2^{( 3,1)}= &(X_1^3+X_2^3)  +\frac{[2]_q}{Q^2}X_1X_2(X_1+X_2)
\label{Omega22131}
\ee
and adjust a couple of coefficients, $\frac{1}{Q}$ and $\frac{1}{Q^2}$ at two places in these formulas.
This is an alternative representation of (\ref{Omega331}), which will use now.

\bigskip

For ${\bf m=2}$, an additional transformation is needed at the level of $\omega$.
Naively
\be
\omega_3^{(3,2)}\! =\left(\omega_{3}^{(3,1)}\right)^2\!\! = B_2^2\,\omega_{2}^{(3,2)}
+ 2 X_3B_2B_1\,\omega_{2}^{(2,1)}\omega_{2}^{(3,2)}  +
\nn \\
+ X_3^2\Big(B_1^2\,\omega_2^{(2,2)}\omega_{2}^{(3,2)}+2B_2\,\omega_2^{(3,3)}\Big)
+ 2X_3^3 B_1\,\omega_2^{(2,1|3,3)}\!
+ \!X_3^4\,\omega_2^{(3,4)}
\label{preprecursorofomega_3^{(3,2)}}
\ee
where the only simplification is the use of property
$\omega_N^{(a,m)}=\big(\omega_N^{(a,1)}\big)^m$.
Our formula contains products of $B_s$, which are deformed under the $Q$-deformation,
therefore it is better to expand it in terms of products of $X_i$:
\be
\omega_3^{(3,2)} = (X_3^8+X_1^4X_2^4)\,\omega_2^{(3,2)}
+ 2X_3(X_3^6+X_1^3X_2^3)\,\omega_2^{(2,1|3,2)}
+ X_3^2(X_3^4+X_1^2X_2^2)\left(2\omega_2^{(3,3)} + \omega_2^{(2,2|3,2)}\right)
+ \nn \\
+ 2X_3^3(X_3^2+X_1X_2)\Big(\omega_2^{(2,1|3,3)} + X_1X_2\,\omega_2^{(2,1|3,2)}\Big)
+ X_3^4\left(\omega_2^{(3,4)} + 2X_1X_2\,\omega_2^{(2,2|3,2)}
+ 2X_1^2X_2^2\,\omega_2^{(3,2)}\right)
\label{precursorofomega_3^{(3,2)}}
\ee
It is natural to {\it assume} that, under deformation, the single $\omega^{(a,m)}$ just changes for $\Omega^{(a,m)}$,
while the products $\omega^{(a_1,m_1|a_2,m_2)}$ deform nontrivially.
Following this logic, it is also natural to try to simplify as many $\omega^{(a_1,m_1|a_2,m_2)}$ as possible.
Actually, such simplifications are readily available for terms with even powers of $X_3$,
so that products of $\omega$ remain only in odd powers (underlined),
while in the double-underlined brackets they can be, and, in fact, they are eliminated:
\be
\omega_3^{(3,2)} = (X_3^8+X_1^4X_2^4)\,\omega_2^{(3,2)}
+ \underline{2X_3(X_3^6+X_1^3X_2^3)\,\omega_2^{(2,1|3,2)}}
+ X_3^2(X_3^4+X_1^2X_2^2)\left(\underline{\underline{3\omega_2^{(3,3)} + X_1X_2\,\omega_2^{(3,2)}}}\right)
+ \nn \\
+ \underline{2X_3^3(X_3^2+X_1X_2)\Big(\omega_2^{(2,1|3,3)} + X_1X_2\,\omega_2^{(2,1|3,2)}\Big)}
+ X_3^4\left(\underline{\underline{\omega_2^{(3,4)} + 2X_1X_2\,\omega_2^{(3,3)}
+ 4X_1^2X_2^2\,\omega_2^{(3,2)}}}\right)
\label{omega_3^{(3,2)}}
\ee
So far these were just identical transformations of the elementary $Q$-independent polynomials.

The point is that the result is deformed {\it straightforwardly}, by making use of our usual technique
for  remaining $\omega_2^{(a_1,m_1|a_2,m_2)}$ and adjusting the deformation factors $Q^c$.

The {\it direct} deformation of (\ref{omega_3^{(3,2)}}) is:
\be
\Omega_3^{(3,2)} = (X_3^8+X_1^4X_2^4)\Omega_2^{(3,2)}
+ \frac{[2]_q}{Q^3}X_3(X_3^6+X_1^3X_2^3)\Omega_2^{(2,1|3,2)}
+ X_3^2(X_3^4+X_1^2X_2^2)\left(\frac{[3]_q}{Q^4}\Omega_2^{(3,3)} + X_1X_2\Omega_2^{(3,2)}\right)
+ \nn
\ee
\be
+ \frac{[2]_q}{Q^3}X_3^3(X_3^2+X_1X_2)\Big(\Omega_2^{(2,1|3,3)} + X_1X_2\Omega_2^{(2,1|3,2)}\Big)
+ X_3^4\left(\Omega_2^{(3,4)} + \frac{[2]_q}{Q^4}X_1X_2\Omega_2^{(3,3)}
+ \frac{[2]_q^2}{Q^3}X_1^2X_2^2\Omega_2^{(3,2)}\right)
\label{Omega_3^{(3,2)}}
\ee
Like (\ref{Omega331p}) this formula still contains bi-$\Omega$ functions,
which we list in the next paragraph.

\paragraph{Examples of $\Omega_2^{(2,1|3,m)}$.}

We already know $\omega_{ 2}^{(2,1|3,1)}$ and $\Omega_{ 2}^{(2,1|3,1)}$ from (\ref{Omega22131}).
Just in the same way
\be
  &
\Omega_2^{(2,1|3,2)} = (X_1^5+X_2^5) + \frac{[3]_q}{Q^4}X_1X_2(X_1^3+X_2^3)+
\left(\frac{1}{Q^6}\frac{[4]_q[3]_q}{[2]_q} - Q^6\right)(X_1X_2)^2 (X_1+X_2)
\ee
\be
&\uparrow
\nn \\
&\omega_2^{(2,1|3,2)}
= (X_1^5+X_2^5) + 3X_1X_2(X_1^3+X_2^3)+
5 (X_1X_2)^2 (X_1+X_2) =
\nn \\
\stackrel{(\ref{omeome})}{=}\
X_2^5 + 3X_1X_2^4+(6-1)X_1^2X_2^3 + &(10-3-2\cdot 1)X_1^3X_2^2+(15-6-2\cdot  3)X_1^4X_2+
(21-10-2\cdot 6 + 2\cdot 1)X_1^5
\nn
\ee
and
\be
  &
\Omega_2^{(2,1|3,3)} = (X_1^7+X_2^7) + \frac{[4]_q}{Q^6}X_1X_2(X_1^5+X_2^5)+
\left(\frac{1}{Q^{10}}\frac{[5]_q[4]_q}{ [2]_q} - Q^8\right)(X_1X_2)^2 (X_1^3+X_2^3)
+ \nn \\
&
+ \left(\frac{1}{Q^{12}}\frac{[6]_q[5]_q[4]_q}{ [3]_q[2]_q}-Q^6[4]_q-Q^3[3]_q\right) X_1^3X_2^3(X_1+X_2)
\ee
Note that the terms with all powers of $X_1$, from 0 to 7 can be obtained from {\it straightforward}
deformation of (\ref{omeome}), but it is simpler to restore the most complicated half of them, from 4 to 7,
just from the $X_1\leftrightarrow X_2$ symmetry.
The main property of these formulas for $\Omega_2^{(2,1|3,m)}$ is that they
continue to satisfy (\ref{peri}) with $\varepsilon^3=1$, so that one can use the same two-step
procedure as in section  \ref{sec_a=2}:  consider the linear combination of
such $(X_1,X_2)$-symmetric $\Omega_2^{(2,1|3,m)}$
and adjust the $X_3$-dependent coefficients in from of them from the requirement of symmetry between
$X_1$ and $X_3$.

Below we will need the next quantities from the same list:
\be
  &
\Omega_2^{(2,1|3,4)} = (X_1^9+X_2^9) + \frac{[5]_q}{Q^8}X_1X_2(X_1^7+X_2^7)+
\left(\frac{1}{Q^{14}}\frac{[6]_q[5]_q}{[2]_q} -Q^{10}\right)(X_1X_2)^2 (X_1^5+X_2^5)
+ \nn \\
& \!\!\!\!\!\!\!\!\!\!\!\!\!\!\!\!\!\!\!\!\!\!
+\left(\frac{1}{Q^{18}} \frac{[7]_q[6]_q[5]_q}{[3]_q[2]_q}-Q^6[5]_q-Q^3[4]_q\right)X_1^3X_2^3(X_1^3+X_2^3)
+ \left(\frac{1}{Q^{20}}\frac{[8]_q[7]_q[6]_q[5]_q}{[4]_q[3]_q[2]_q} - Q^4\frac{[6]_q[5]_q}{[2]_q} -\frac{1}{Q^2}{[5]_q[4]_q}\right)
X_1^4X_2^4(X_1+X_2)\nn\\
\ee
\be
  &
\Omega_2^{(2,1|3,5)} = (X_1^{11}+X_2^{11}) + \frac{[6]_q}{Q^{10}}X_1X_2(X_1^9+X_2^9)+
\left(\frac{1}{Q^{18}}\frac{[7]_q[6]_q}{ [2]_q} -Q^{12}\right)(X_1X_2)^2 (X_1^7+X_2^7)
+ \nn \\
&\!\!\!\!\!\!\!\!\!\!\!\!\!\!\!\!\!\!\!\!\!\!+ \left(\frac{1}{Q^{24}}\frac{[8]_q[7]_q[6]_q}{[3]_q[2]_q} - Q^6[6]_q-Q^3[5]_q\right) X_1^3X_2^3 (X_1^5+X_2^5)
+ \left(\frac{1}{Q^{28}}\frac{[9]_q[8]_q[7]_q[6]_q}{[4]_q[3]_q[2]_q} - Q^2\frac{[7]_q[6]_q}{[2]_q}-\frac{1}{Q^4}[6]_q[5]_q\right)
X_1^4X_2^4 (X_1^3+X_2^3)
+\!\!\!\!\!\! \nn \\
&+\left(  \frac{1}{Q^{30}}\frac{[10]_q[9]_q[8]_q[7]_q[6]_q}{[5]_q[4]_q[3]_q[2]_q}-
 \frac{[8]_q[7]_q[6]_q}{[3]_q[2]_q} - \frac{1}{Q^9}[5]_q\frac{[7]_q[6]_q}{[2]_q} + Q^{15} [5]_q \right)  X_1^5X_2^5 (X_1+X_2)
\ee
In general,
\be
\Omega_2^{(2,1|3,m)} = \sum_{i-0}^{2m+1} X_1^i X_2^{2m+1-i} \!\!\!
\sum_{k_2=0}^{{\rm min}\left(m,\left[\frac{i }{3}\right]\right)} \ \
\sum_{k_1=0}^{{\rm min}\left(1,\left[\frac{i-3k2}{2}\right]\right)}  \!\!\!
(-)^{k_1+k_2} Q^\mu \ \frac{[m]_q!}{[k_2]_q![m-k_2]_q!} \frac{[m+i-2k_1-3k_2]_q!}{[m]_q![i-2k_1-3k_2]_q!}
\label{Ome2^{(2,1|3, )}}
\label{OmeOme2213}
\ee
where $\mu = -2im+i(i-1)+6m(k_1+k_2)-3(i-2)k_2 -6k_1k_2+\frac{3}{2}k_2(k_2-1)$.

Further we can increase $m$ in the first factor, e.g.
\be
\Omega_2^{(2,2|3,1)} = (X_1^4+X_2^4) + \frac{[3]_q}{Q^3}X_1X_2(X_1^2+X_2^2)+
\left(\frac{1}{Q^4}\frac{[4]_q[3]_q}{[2]_q} -[2]_q Q^5\right)(X_1X_2)^2
\nn \\ \nn \\
\Omega_2^{(2,2|3,2)} = (X_1^6+X_2^6) + \frac{[4]_q}{Q^5}X_1X_2(X_1^4+X_2^4)+
\left(\frac{1}{Q^8}\frac{[5]_q[4]_q}{[2]_q} -[2]_q Q^7\right)(X_1X_2)^2 (X_1^2+X_2^2)
+ \nn \\
+ \left(\frac{1}{Q^9}\frac{[6]_q[5]_q[4]_q}{[3]_q[2]_q}  -  Q^3[2]_q \boxed{\Big(Q^3[3]_q+1\Big)}-Q^9[2]_q\right)  X_1^3X_2^3
= \Omega_2^{(3,3)}+Q^4X_1X_2\Omega_2^{(3,2)}
\label{preOmega2223}
\ee
Notice the coefficient in the box, which substitutes the naive $[4]_q$.
Such modification is needed for $\Omega_2^{(2,2|3,2)}$ to satisfy the equation (\ref{peri}) at $a=3$ and $m=2$.

At this point it deserves explaining the difference between (\ref{preprecursorofomega_3^{(3,2)}}),
(\ref{precursorofomega_3^{(3,2)}}) and (\ref{omega_3^{(3,2)}}) in more detail:
(\ref{preprecursorofomega_3^{(3,2)}}) contains squares of sums
like $B_2^2 = (X_3^4+X_1^2X_2^2)^2 = X_3^8+2X_1^2X_2^2X_3^4 + X_1^4X_2^4$,
and deformation naturally changes coefficient $2$ for $[2]_q$, thus it does not leave
the entire $B_2^2$ intact.
This effect is taken into account by transition from  (\ref{preprecursorofomega_3^{(3,2)}}) to
(\ref{precursorofomega_3^{(3,2)}}).
But after that, a more subtle fact is revealed by the very last identity in (\ref{preOmega2223}):
it turns out that $\Omega_2^{(2,2|3,2)}$ is not an independent quantity, and when its expression
through $\Omega_2^{(3,\ )}$ is substituted into the naively deformed (\ref{precursorofomega_3^{(3,2)}}),
a new factor of $2$ appears, that is, at $2[2]_qX_1^2X_2^2\Omega_2^{(3,2)}$.
Naturally, it is instead deformed to $[2]_q^2$, with an additional factor of $Q^{-3}$.
This effect is already seen at the non-deformed level (\ref{omega_3^{(3,2)}}),
and is thus a necessary part of precooking an appropriate expression for the {\it direct} deformation,
which consists of replacing numbers for quantum numbers and insertion of powers of $Q$.
This is already a limited freedom, which is easier to handle with the help of
(\ref{peri}) and the symmetry requirements before the final check
that the answer indeed provides an eigenfunction.
$\Omega_3^{(3,2)}$ in (\ref{Omega_3^{(3,2)}}) does.

\subsection{$m=3$ at $a=3$
\label{m3ata3}}

Now we are prepared for the next attempt, that at ${\bf m=3}$.
This is one of the highest examples available by directly solving the periodicity equations. The answer reads

\be
\Omega_3^{(3,3)} = (X_3^{12}+X_1^6X_2^6)\Omega_2^{(3,3)}
+ \frac{[3]_q}{Q^5}X_3(X_3^{10}+X_1^5X_2^5)\Omega_2^{(2,1|3,3)}
+ \nn \\
+ X_3^2 (X_3^8+X_1^4X_2^4)\left(\frac{1}{Q^8}\frac{[4]_q[3]_q}{[2]_q}\Omega_2^{(3,4)}
+ \frac{[3]_q}{Q^3}X_1X_2\Omega_2^{(3,3)}\right)
+ \nn \\
 + X_3^3(X_3^{6}+X_1^3X_2^3)\left(\frac{\boxed{[4]_q+Q^6[3]_q}}{Q^9}\Omega_2^{(2,1|3,4)}
+\frac{[2]_q^2}{Q^5}X_1X_2\Omega_2^{(2,1|3,3)}\right)
+ \nn \\
+ X_3^4(X_3^4+X_1^2X_2^2)\left(\frac{1}{Q^8}\frac{[4]_q[3]_q}{[2]_q}\Omega_2^{(3,5)}
+\frac{[3]_q^2}{Q^8}X_1X_2\Omega_2^{(3,4)}+\frac{[3]_q^2}{Q^6}X_1^2X_2^2\Omega_2^{(3,3)}\right)
+ \nn \\
 + \frac{[3]_q}{Q^5}X_3^5(X_3^2+X_1X_2)\left(\Omega_2^{(2,1|3,5)}+\frac{[3]_q}{Q^4}X_1X_2\Omega_2^{(2,1|3,4)}
+ \frac{[3]_q}{Q^3}X_1^2X_2^2\Omega_2^{(2,1|3,3)}\right)
+ \nn \\
+X_3^6\left(\Omega_2^{(3,6)} + \frac{[3]_q[2]_q}{Q^8}X_1X_2\Omega_2^{(3,5)}
+ \boxed{\frac{Q^3+2}{Q^8}}\cdot\frac{[4]_q[3]_q}{[2]_q}X_1^2X_2^2\Omega_2^{(3,4)}
+\frac{\boxed{2}[3]_q}{Q^3}X_1^3X_2^3\Omega_2^{(3,3)}\right)
\label{Omega333}
\ee
Three strange coefficients appeared in this formula,
they are marked by boxes (note that $2$ in the last box is not quantum).

The three preparatory steps at the $Q\to 1$ level, like (\ref{preprecursorofomega_3^{(3,2)}}),
(\ref{precursorofomega_3^{(3,2)}}) and (\ref{omega_3^{(3,2)}}) this time are:
\be
\omega_{3}^{(3,1)} = B_2\,\omega_{2}^{(3,1)}
&+ X_3B_1\,\omega_{2}^{(2,1)}\omega_{2}^{(3,1)}
&+ X_3^2\,\omega_{2}^{(3,2)}
\nn \\ \nn \\
&\downarrow
\nn
\ee
\be
\omega_3^{(3,3)}\! =\left(\omega_{3}^{(3,1)}\right)^3\!\! = B_2^3\,\omega_{2}^{(3,3)}
+ 3 X_3B_2^2B_1\,\omega_{2}^{(2,1)}\omega_{2}^{(3,3)}
+ 3 X_3^2B_2B_1^2\,\omega_{2}^{(2,2)}\omega_{2}^{(3,3)}
 + X_3^3B_1^3\,\omega_2^{(2,3)}\omega_2^{(3,3)} +
\ee
\vspace{-0.7cm}
\be
+ 3X_3^2B_2^2\,\omega_2^{(3,4)}
 + 6X_3^3B_2B_1\,\omega_2^{(2,1)}\omega_2^{(3,4)}
+3X_3^4B_1^2\,\omega_2^{(2,2)}\omega_2^{(3,4)}
+ 3X_3^4B_2\,\omega_2^{(3,5)}+3X_3^5B_1\,\omega_2^{(2,1)}\omega_2^{(3,5)}
+ \!X_3^6\,\omega_2^{(3,6)}
\nn
\label{preprecursorofomega_3^{(3,3)}}
\ee
The analogue of (\ref{precursorofomega_3^{(3,2)}}),
where all higher powers and products of $B_i$ are eliminated, is
\be
\omega_3^{(3,3)} = B_6\omega_2^{(3,3)} + 3X_3B_5\,\omega_2^{(2,1|3,3)}
+ 3X_3^2B_4 \Big(\omega_2^{(3,4)}+\omega_2^{(2,2|3,3)}\Big)
+ \nn \\
+X_3^3 B_3\Big(\omega_2^{(2,3|3,3)}+6\omega_2^{(2,1|3,4)} +3X_1X_2\,\omega_2^{(2,1|3,3)}\Big)
+ \nn \\
+3X_3^4 B_2\Big( \omega_2^{(3,5)}+\omega_2^{(2,2|3,4)}
+ 2X_1X_2\,\omega_2^{(2,2|3,3)}+X_1^2X_2^2\,\omega_2^{(3,3)}\Big)
+ \nn \\
+ 3X_3^5B_1\left(\omega_2^{(2,2|3,5)}+X_1X_2\Big(2\omega_2^{(2,1|3,4)}+\omega_2^{(2,3|3,3)}\Big)
+ 2X_1^2X_2^2\,\omega_2^{(2,2|3,3)} \right)
+ \nn \\
+X_3^6 \left(\omega_2^{(3,6)}+ 6X_1X_2\,\omega_2^{(2,2|3,4)}
+ 6X_1^2X_2^2\Big(\omega_2^{(3,4)}+\omega_2^{(2,2|3,3)}\Big) \right)
\label{precursorofomega_3^{(3,3)}}
\ee
The final step is transformation of bi-$\omega_2$ functions:
\be
\omega_3^{(3,3)} = B_6\,\omega_2^{(3,3)} + 3X_3B_5\,\omega_2^{(2,1|3,3)}
+  3X_3^2B_4 \Big(2\omega_2^{(3,4)}+ X_1X_2\,\omega_2^{(3,3)}\Big)
+ \nn \\
+X_3^3 B_3\Big(\boxed{7}\omega_2^{(2,1|3,4)}+4X_1X_2\,\omega_2^{(2,1|3,3)}\Big)
+ \nn \\
+3X_3^4 B_2\Big( \omega_2^{(3,5)}+3X_1X_2\,\omega_2^{(3,4)}+3X_1^2X_2^2\,\omega_2^{(3,3)}\Big)
+ \nn \\
+ 3X_3^5B_1\Big(\omega_2^{(2,1|3,5)}+X_1X_2\, \omega_2^{(2,1|3,4)}+X_1^2X_2^2\,\omega_2^{(2,1|3,3)}  \Big)
+ \nn \\
+X_3^6 \Big(\omega_2^{(3,6)}+ 6X_1X_2\,\omega_2^{(3,5)}
+ \boxed{18}X_1^2X_2^2\,\omega_2^{(3,4)}+\boxed{6}X_1^3X_2^3\,\omega_2^{(3,3)}\Big)
\label{omega_3^{(3,3)}}
\ee
Like in (\ref{omega_3^{(3,2)}}), it turns possible to eliminates all bi-$\omega$
in the terms with even powers of $X_3$, and this time also possible is  certain simplification
and ordering in the odd-power terms.
Important is also elimination of all bi-$\omega_2$, except for the set $(2,1|3,m)$
where we know the deformation (\ref{Ome2^{(2,1|3, )}}) in full generality.
We emphasize that at $Q=1$ all these are equivalent formulas, all are equally correct.
But {\it direct} deformation,
term-by-term conversion to solution (\ref{Omega333}) of eigenvalue equations
is possible only for (\ref{omega_3^{(3,3)}}).
Nontrivially deformed are the three terms in the boxes.

\bigskip

In general, the analogue of (\ref{omega_3^{(3,2)}}) and (\ref{omega_3^{(3,3)}}) is
\be
\omega_3^{(3,m)} = \sum_{i=0}^m X_3^{2i}B_{2m-2i}
\cdot \sum_{j=0}^i \mu^{(m)}_{i,j}X_1^jX_2^j\,\omega_2^{(3,m+i-j)}
+ \sum_{i=0}^{m-1} X_3^{2i+1}B_{2m-2i-1}
\cdot\sum_{j=0}^i \nu^{(m)}_{i,j}X_1^jX_2^j\,\omega_2^{(2,1|3,m+i-j)}
\label{decoomega33}
\ee
and the coefficients may look somewhat ugly:
\be
\mu^{(m)}_{0,0} = 1, \ \ \ \ \ \ \ \  \nu^{(m)}_{0,0}=m,
\nn \\
\mu^{(m)}_{1,0} = \frac{m(m+1)}{2}, \ \ \ \mu^{(m)}_{1,1} = \frac{m(m-1)}{2},
\nn \\
 \nu^{(m)}_{1,0}=\frac{m(m-1)(m+4)}{6}, \ \ \nu^{(m)}_{1,1}= \frac{m(m^2-3m+8)}{6},
\nn \\
\ \ \mu^{(m)}_{2,0} = \frac{m(m-1)(m^2+7m-6)}{24}, \ \ \mu^{(m)}_{2,1} = \frac{m(m-1)(m^2+m+6)}{12},
\ \ \mu^{(m)}_{2,2} = \frac{m(m^3-6m^2+35m-6)}{24},
\nn \\
\nu^{(m)}_{2,0} = \frac{(m+1)m(m-1)(m-2)(m+12)}{120}, \ \
\nn\\
\mu^{(m)}_{3,0} = \frac{m(m-1)(m-2)(m^3 + 18m^2 + 17m - 120)}{6!}, \ \
\nn \\
\nu^{(m)}_{3,0} = \frac{(m+1)m(m-1)(m-2)(m^3 + 69m^2 -178m - 120)}{7!\cdot 2}, \ \
\nn \\
\mu^{(m)}_{4,0} = \frac{m(m-1)(m-2)(m-3)(m^4+34m^3 + 179m^2 -694m - 840)}{8!}, \ \
\nn \\  \nn \\
\ldots
\nn \\  \nn \\
\mu^{(m)}_{m,0}=1,\ \ \mu^{(m)}_{m,1}=m(m-1),\ \ \mu^{(m)}_{m,2}=\frac{m(m-1)(m^2-m+6)}{4},
\ \ \ \ldots
\ee
These coefficients can actually be obtained directly as functions of $m$
as an iterative solution of the system
\be
(1+zx+z^2x^2)^m(1+zy+z^2y^2)^m = \sum_{i=0}^m z^{2i}\mu^{(m)}_{2i,j} \sum_{j=0}^i (xy)^j(x^2+xy+y^2)^{i-j}
+ \nn \\
+ \sum_{i=0}^{m-1} z^{2i+1}\nu^{(m)}_{2i+1,j} \sum_{j=0}^i (x+y) (xy)^j(x^2+xy+y^2)^{i-j}
\ee
treated as a series, say, in $z$ and $y$.
It arises by substitution of  $X_1=zxX_3$ and $X_2=yzX_3$ into (\ref{decoomega33}).
The factors  $1+(z^2xy)^{2m-2i}$ and $1+(z^2xy)^{2m-2i-1}$ at the r.h.s. can and are omitted, because they
do not contribute to terms with $z^{m'}$ with $m'\leq m$, which are analyzed in order
to extract $\mu^{(m')}_{ij}$ and $\nu^{(m')}_{ij}$.
There is only one $\mu^{(m)}_{ij}$ or $\nu^{(m')}_{ij}$ contributing at a given order,
thus solving the linear equations for them is just trivial: the matrix to be inverted is diagonal.

The same simplicity persists under the deformation,
{\bf if} one chooses the deformation of particular $\omega_2$'s satisfying the periodicity equations (\ref{peri}).
Then the deformed coefficients $\mu$ and $\nu$
are defined from the linear condition of symmetry between $X_3$ and $X_{1,2}$,
and again the variables are separated, at a given order in $X's$, only a single coefficient contributes
and can be easily extracted.
This can be done for every particular $m$, and then one can find an interpolating formula,
generalizing the non-deformed expressions.

However, a {\it  direct} deformation prescription  for  $\{\mu,\nu\}$
is not quite apparent.
The factors $(m-k)$ with $k\geq 0$ reflect just the fact that $i\leq m$ and are not very interesting.
They naturally survive after deformation.
But non-factorized coefficients need a special deformation prescription:
the  boxes in    (\ref{omega_3^{(3,3)}}) were just the first examples
of this problem.
Sometime, it is actually not too difficult to handle, but the general procedure remains unclear.

An appropriate deformation consistent with (\ref{peri}) is
\be\label{67}
\Omega_3^{(3,m)} = \sum_{i=0}^m X_3^{2i}B_{2m-2i}
\cdot \sum_{j=0}^i \mu^{(m)}_{i,j}X_1^jX_2^j\,\Omega_2^{(3,m+i-j)}
+ \sum_{i=0}^{m-1} X_3^{2i+1}B_{2m-2i-1}
\cdot\sum_{j=0}^i \nu^{(m)}_{i,j}X_1^jX_2^j\,\Omega_2^{(2,1|3,m+i-j)}
\ee
with $\Omega$'s from (\ref{OmegaN2Q}) and (\ref{OmeOme2213})
and with the coefficients
\be
\mu^{(m)}_{0,0} = 1,
\nn \\ \nn \\
\nu^{(m)}_{0,0}=\frac{[m]_q}{Q^{2m-1}},
\nn \\ \nn \\
\mu^{(m)}_{1,0} = \frac{1}{Q^{4(m-1)}}\cdot \frac{[m]_q[m+1]_q}{[2]_q}, \ \ \
\mu^{(m)}_{1,1} = \frac{1}{Q^{3(m-2)}}\cdot \frac{[m-1]_q[m]_q}{[2]_q},
\nn \\  \nn \\
\nu^{(m)}_{1,0} = \frac{1}{Q^{3(2m-3)}}\cdot \frac{[m+1]_q!}{[3]_q![m-2]_q!}
+ \frac{1}{Q^{3(m-2)}}\cdot \frac{[m]_q!}{[2]_q![m-2]_q!}, \ \ \ \ \
\nu^{(m)}_{1,1} = \frac{1}{Q^{5m-13}}\cdot \frac{[m]_q!}{[3]_q![m-3]_q!}
+ \frac{1}{Q^{2m-1}}\cdot [m]_q,
\nn \\ \nn \\
\mu^{(m)}_{2,0} = \frac{1}{Q^{8(m-2)}}\cdot \frac{[m+2]_q!}{[4]_q![m-2]_q!}
+ \frac{1}{Q^{5m-13}}\cdot \frac{[m]_q!}{[3]_q![m-3]_q!},
\ \ \ \nn \\
\mu^{(m)}_{2,1} = \frac{1}{Q^{7m-13}}\frac{[m+1]_q!}{[4]_q![m-2]_q!}\Big([m+2]_q+(Q^9+Q^6-1)[m-2]_q\Big)
+ \frac{1}{Q^{4(m-1)}}\frac{[m]_q!}{[2]_q![m-2]_q!}
= \nn \\
= \frac{1}{Q^{7m-19}}\frac{[m+1]_q!}{[4]_q[3]_q[m-3]_q!}
+\frac{1}{Q^{4m-7}} \frac{[m+1]_q!}{[3]_q![m-2]_q!} + \frac{1}{Q^{4m-4}}\frac{[m]_q!}{[2]_q![m-2]_q!}
\ \ \ \ \ \ \ \
\nn\\
\mu^{(m)}_{2,2} = \frac{1}{Q^{6(m-4)}}\cdot \frac{[m ]_q!}{[4]_q![m-4]_q!}
+ \frac{1}{Q^{3(m-1)}}\cdot [m]_q^2,
\nn\\ \nn \\
\nu^{(m)}_{2,0} = \frac{1}{Q^{2(2m-5)}}\cdot \frac{[m+2]_q!}{[5]_q![m-3]_q!}
+ \frac{[2]_q}{Q^{7m-19}}\cdot \frac{[m+1]_q!}{[4]_q![m-3]_q!},
\ \ \
\nn \\ \nn \\
\ldots
\nn \\ \nn\\
\boxed{
\mu^{(m)}_{m-i,0} = \mu^{(m)}_{i,0}, \ \ \ \ \ \nu^{(m)}_{m-1-i,0} = \nu^{(m)}_{i,0}
}
\ee
In many respects, $\nu$ and $\mu$ form a common series.

For some concrete results for $\Omega_3^{(3,4)}$, $\Omega_3^{(3,5)}$ and $\Omega_3^{3,6}$
in the maple format, see file Ome3.txt attached to this submission.

\section{Excitations}

Now we are ready to discuss the structure of excitations over $\Omega_N^{(a,m)}$. In fact, these results were already presented in \cite{MMPns}, but here we formulate them in a different form.

\subsection{Trivial example}

We start with a special set of eigenfunctions,
\be
\Psi^{(a,m)}_{[\underbrace{0,0,\ldots,0}_{N-k},\underbrace{1,\ldots 1}_k]} =
\left(\prod_{i=N-k}^k X_i\right)\cdot \Omega_N^{(a,m)}(X_1,\ldots,X_{N-k-1},
QX_{N-k},\ldots, QX_N)
\ee
which is very simple. In this case,  in (\ref{Zdeco0}) $E_\alpha^{\sigma}[\vec x]=\delta_{\alpha\sigma}$, and $\Psi^{(a,m)}_{[0,0,\ldots,0,1,\ldots 1]}=\Xi^{(a,m)}_{[0,0,\ldots,0,1,\ldots 1]}$.

\subsection{How it works: obtaining $\Psi_{[0,1,0]}^{(2,m)}$}

Turning to non-trivial excitations, we begin with the example of
$\Psi_{[0,1,0]}^{(2,m)}$, which illustrates the problems to solve and the ways
to do it along the same lines as we did for $\Omega$.

As usual, we begin with the case of $Q=1$.
Then the level one excitation with unity at the place $s$ is
\be
\psi^{(a,m)}_{[0,\ldots,0,1_s,0,\ldots 0]}
= \left((sm-1)X_s + m \sum_{s'=s+1}^N X_{s'}\right)\omega_N^{(a,m)}
\ee
where $\omega_N^{(a,m)}$ is given by (\ref{omega}).
As we did in (\ref{e2L0alt}), we now switch to another, ``triangular'' basis:
\be
\psi^{(a,m)}_{[0,\ldots,0,1_s,0,\ldots 0]}
= \left\{(sm-1)(X_s-X_{s+1}) + \Big(s+1)m-1\Big)(X_{s+1}-X_{s+2}) + \ldots
+ \nn \right. \\ \left.
+ \Big((N-1)s-1\Big)(X_{N-1}-X_N) + (Ns-1) X_N
\right\} \omega_N^{(a,m)}
\ee
Coming closer to our concrete examples, at $N=2$,
\be
\psi^{(a,m)}_{[0,1]} &=&  X_2 \cdot \left(\frac{X_1^a-X_2^2}{X_1-X_2}\right)^m\nn\\
\psi^{(a,m)}_{[1,0]} &=& \Big\{  (m-1)(X_1-X_2) + (2m-1)X_2\Big\}\left(\frac{X_1^a-X_2^2}{X_1-X_2}\right)^m
\label{z2m}
\ee
Overall normalization does not matter, therefore we change it freely.
The deformation of (\ref{z2m}) is simple, and not very surprising after (\ref{ZL0}) and (\ref{Z0L}):
\be
\Psi^{(2,m)}_{[0,1]} &=&   X_2 \Omega_2^{2,m)}(X_1,QX_2)\nn\\
\Psi^{(2,m)}_{[1,0]} &=& Q^{(2a-1)m}   [m-1]_q (Q^{-am}X_1^a-X_2^a)\cdot \Omega_2^{(2,m-1)}(X_1,X_2) + [2m-1]_qX_2 \Omega_2^{2,m)}(X_1,QX_2)
\label{Z2m}
\ee
The main thing  happens in the first term, where $X_1-X_2$ cancels one power in the denominator.

At $N=3$ case, basically the same thing happens, together with the standard deformation of sum powers
like $(A+B)^2 \longrightarrow (A+QB)(A+Q^{-1}B) = A^2+[2]_qAB+B^2$ etc.
In the simplest case of $a=2$, the non-deformed formulas are
\be
\psi^{(2,m)}_{[0,0,1]} &=&   X_3 \Big((X_1+X_2)(X_1+X_3)(X_2+X_3)\Big)^m\nn\\
\psi^{(2,m)}_{[0,1,0]} &=& \Big\{  (2m-1)(X_2-X_3) + (3m-1)X_3\Big\}\Big((X_1+X_2)(X_1+X_3)(X_2+X_3)\Big)^m  \nn \\
\psi^{(2,m)}_{[1,0,0]} &=& \Big\{(m-1)(X_1-X_2) + (2m-1)(X_2-X_3) + (3m-1)X_3\Big\}\Big((X_1+X_2)(X_1+X_3)(X_2+X_3)\Big)^m
\label{Z3m}
\ee
As we already know, the first line deforms to
\be
\Psi^{(a,m)}_{[0,0,1]} =   X_3 \cdot \Omega_3^{(a,m)}(X_1,X_2,QX_3)
\ee
actually, for any $a$.
Deformation of powers is hidden in formulas for $\Omega_3^{(a,m)}$.
However, at the second line they should be explicitly revealed.
First, similarly to (\ref{Z2m}),
\be
\Psi^{(2,m)}_{[0,1,0]} = [2m-1]_q Q^{m(m+1)} (Q^{-2m }X_2^2-X_3^2)\cdot W_{3}^{(2,m)}(X_1,X_2,X_3)
 + [3m-1]_q Q^{m(m-1)} X_3 \Omega_3^{(2,m)}(X_1,X_2,QX_3)
\label{Z2vsW32}
\ee
with some yet unknown polynomial $W_{3}^{(2,m)}(X_1,X_2,X_3)$, which one needs to evaluate.

The factor $X_2+X_3$ in $\omega_3^{(2,1)}$ gets combined with $(X_2-X_3)$
and provides
$X_2^2-X_3^2$, which becomes $(Q^{-2m }X_2^2-X_3^2)$ after deformation.
The quantity  $W_3^{(2,m)}$ is now a deformation of $(X_1+X_2)(X_1+X_3) \omega_3^{(2,m-1)}$,
rather than just $\omega_2^{(2,m-1)}$ as it was in the case of $N=2$.
There the answer was easy to guess: $\Omega_2^{(2,m-1)}$, probably with some power of $Q$.
Now we should be more careful.

The first case of $m=1$ should be just taken as a fact:
\be
W_3^{(2,1)} = {\rm deform}\Big\{(X_1+X_2)(X_1+X_3)\Big\} =  QX_3(X_1+X_2)+ X_1^2 + QX_1X_2
\ee
The second case is already more informative:
\be
W_3^{(2,2)} = {\rm deform}\Big\{(X_1+X_2)(X_1+X_3)\omega_3^{(2,1)}\Big\}
\ \stackrel{(\ref{Omega32})}{=}
{\rm deform}\Big\{W^{(2,1)}_3\Big( (X_3^2+X_1X_2)\Omega_2^{(2,1)} + X_3\Omega_2^{(2,2)}\Big)\Big\}
= \nn \\
= {\rm deform}\left\{\Big(X_1^2 + QX_1X_2+ QX_3(X_1+X_2)  \Big)\Big((X_3^2+X_1X_2)(X_1+X_2) + X_3(X_1+X_2)^2\Big)\right\}
\ee
The main point is now combining of $(X_1+X_2)$ from two brackets into higher powers,
which leads to
\be
(X_1^2+ X_1X_2)(X_3^2+X_1X_2)(X_1+X_2) +  X_3(X_1^2+ X_1X_2)(X_1^2+[2]_qX_2^2+X_2^2)
+ \nn \\
X_3(X_3^2+X_1X_2)(X_1^2+[2]_qX_1X_2+X_2^2)
 +   X_3^2(X_1^3+[3]_qX_1^2X_2+[3]_qX_1X_2^2+X_3^3)
\ee
The true answer differs by adding a few powers of $Q$ at different places (underlined):
\be
W_3^{(2,2)} =
(X_1^2+\underline{Q^2}X_1X_2)(X_3^2+X_1X_2)(X_1+X_2) + X_3(X_1^2+QX_1X_2)(X_1^2+[2]_qX_2^2+X_2^2)
+ \nn \\
+ \underline{Q^2}X_3(X_3^2+\underline{Q}X_1X_2)(X_1^2+[2]_qX_1X_2+X_2^2) +
\underline{Q^2}X_3^2(X_1^3+[3]_qX_1^2X_2+[3]_qX_1X_2^2+X_3^3)
\ee

Now we proceed a little more systematically.
We begin with the $Q=1$ expression for $W$, expanded in powers of $X_3$ and then in powers of $X_1+X_2$
with positive coefficients which are monomials in $X_1$ and $X_2$:
\be
w_3^{(2,1)} = X_1(X_1+X_2)+X_3(X_1+X_2),
\nn \\ \nn \\
w_3^{(2,2)} = X_1^2X_2(X_1+X_2)^2 +X_3X_1X_2(X_1+X_2)^2+ X_3X_1(X_1+X_2)^3 + \nn \\
+X_3^2X_1(X_1+X_2)^2 + X_3^2(X_1+X_2)^3 + X_3^3(X_1+X_2)^2,
\nn \\ \nn \\
w_3^{(2,3)} = X_1^3X_2^2(X_1+X_2)^3  +X_3X_1^2X_2^2(X_1+X_2)^3 + 2X_3X_1^2X_2(X_1+X_2)^4 + \nn \\
+2 X_3^2X_1^2X_2(X_1+X_2)^3 +2 X_3^2X_1X_2(X_1+X_2)^4+X_3^2X_1(X_1+X_2)^5 + \nn \\
+2X_3^3X_1X_2(X_1+X_2)^3+2X_3^3X_1(X_1+X_2)^4 +X_3^3(X_1+X_2)^5+ \nn \\
+X_3^4X_1(X_1+X_2)^3 +2X_3^4(X_1+X_2)^4+ X_3^5(X_1+X_2)^3,
\nn \\ \nn \\
\ldots
\ee
Of course, the minimal power in $X_1+X_2$ is $m$, but for the purposes of the {\it proper} deformation
we need a more detailed separation in monomials, which produces also higher powers of $X_1+X_2$.
For an arbitrary $m$,
\be
\psi_{[0,1,0]}^{(2,m)} = ((2m-1)X_2+mX_3)\omega_3^{(2,m)} = \nn \\
\!=(2m-1)\underbrace{(X_2-X_3)\Big((X_1+X_2) (X_1+X_2)(X_1+X_3)\Big)^m}_{(X_2^2-X_3^2)w_3^{(2,m)}}
\!\!+ (3m-1)X_3\Big((X_1+X_2) (X_1+X_2)(X_1+X_3)\Big)^m
\ee
and
\be
w_3^{(2,m)} =  (X_2+X_3)^{m-1}\Big((X_1+X_2)(X_1+X_3)\Big)^m = \nn \\
= \sum_{j=0}^{2m-1} X_3^j \left(\sum_{i=0}^j
\frac{u_{i,j}\cdot  (m-1)!}{{\rm floor}\left(\frac{j+i+1}{2}\right)!\left(m-{\rm floor}\left(\frac{j+i+2}{2}\right)\right)!}
(X_1+X_2)^{m+i}
X_1^{m-{\rm floor}\left(\frac{j+i+1}{2}\right)} X_2^{m-{\rm floor}\left(\frac{j+i+2}{2}\right)}\right)
\label{w32m}
\ee
with the coefficients
\be
u[i,j] := \frac{1}{i!} \prod_{k=1}^i {\rm floor}\left(\frac{j-i}{2}+k\right)
\ee
These coefficients are deformed as
\be
U[i,j] := \frac{1}{[i]_s!} \prod_{k=1}^i \left[{\rm floor}\left(\frac{j-i}{2}+k\right)\right]_s
\ee

Now deformation involves three operations: replacing numbers by quantum numbers,
replacing powers of sums $X_1+X_2$ by Pochhammer-like expressions, and
insertion of powers of $Q$ in front of monomials:
$[X_1+X_2]_k^m:=\prod_{i=0}^{m-1}(X_1+Q^{k+2i}X_2)$.
Deformation of sums also involves a discrete freedom, a choice of the ``initial'' point,
in a sense similar to the last operation. Then,
\be
W_3^{(2,1)} =  X_1(X_1+\underline{Q}X_2) +\underline{Q}X_3(X_1+X_2)
=  X_1[X_1+ X_2]_1 +\underline{Q}X_3[X_1+X_2]_0,
\nn \\
W_3^{(2,2)} = X_1^2X_2[X_1+X_2]_0^2
+ \underline{Q}X_3X_1X_2[X_1+X_2]^2_{-1} + X_3X_1[X_1+X_2]^3_{-1} + \nn \\
+X_3^2X_1[X_1+X_2]_0^2 + \underline{Q^2}X_3^2[X_1+X_2]_{-2}^3 + \underline{Q^2}X_3^3[X_1+X_2]_{-1}^2,
\nn \\
W_3^{(2,3)} = X_1^3X_2^2[X_1+X_2]^3_{-1}  +\underline{Q}X_3X_1^2X_2^2[X_1+X_2]^3_{-2}
+ [2]_sX_3X_1^2X_2[X_1+X_2]^4_{-2} + \nn \\
+[2]_s X_3^2X_1^2X_2[X_1+X_2]^3_{-1} +[2]_s\underline{Q^2} X_3^2X_1X_2[X_1+X_2]^4_{-3}
+X_3^2X_1[X_1+X_2]_{-3}^5  + \nn \\
+[2]\underline{Q^2}X_3^3X_1X_2[X_1+X_2]^3_{-2}   +[2]_sX_3^3X_1[X_1+X_2]^4_{-2}
+\underline{Q^3}X_3^3[X_1+X_2]^5_{-4}+ \nn \\
+X_3^4X_1[X_1+X_2]^3_{-1} +[2]_s\underline{Q^3}X_3^4[X_1+X_2]^4_{-3}+ \underline{Q^3}X_3^5[X_1+X_2]^3_{-2},
\nn \\
\ldots
\label{W32exa}
\ee

It is now clear that (\ref{w32m}) provides an appropriate representation for $w$
for the {\it proper} deformation.
To understand the deformation rules, i.e. the powers of $Q$ and the shifts in Pochhammer products,
it is useful to list them in the table:
{\footnotesize
\be
\!\!\!\!\!\!
\begin{array}{c||c||c|c||c|c|c||c|c|c|c||c|c|c|c|c||c|c }
\phantom{.}\j  &  0 & 1 && 2 &&& 3 &&&& 4 &&&&& 5 &  \\
m \backslash i & 0 & 0 & 1 & 0 & 1 & 2 & 0 &1&2&3 & 0&1&2&3&4& 0&\ldots   \\
\hline
1  &  \underline{1,1}  & \underline{Q,0} &  &&&&  &&&& &&&& & \\
2 &  1,0 & Q,-1 & \underline{\underline{\underline{1,-1}}}& \underline{\underline{1,0}}
& \underline{\underline{Q^2,-2}} & & \underline{Q^2, -1} &  &  &&&&  &&&\\
3 & 1,-1 & Q,-2 & 1,-2& 1,-1& Q^2,-3&1,-3
& Q^2,-2& \underline{\underline{\underline{1,-2}}}&\underline{\underline{\underline{Q^3,-4}}}& & \underline{\underline{1,-1}}
&\underline{\underline{Q^3,-3}} &&&&\underline{Q^3,-2}   \\
\end{array}
\nn\
\ee
}

\noindent
The table is made from the examples in (\ref{W32exa}).
Many rules are already clear.
For example, in each column the first item (power of $Q$) does not change,
while the second item (shift in the Pochhammer product) increases by one with each line.
Clear are also some ``initial conditions'', the first items in the column:
they are underlined, and the number of underlines counts the distance $2m-j$ from the right.
Undefined for $m=4$ remain then just two items in italics, but they are already easy to guess from
the structure of boxes with individual $j$:
clearly, the entries switch between $1$ and increasing power of $Q$ at the first position,
and the entry at the second position jumps by $-2$.
From these observations we can guess the next lines for higher $m$ as well:

{\footnotesize
\be
\!\!\!\!\!\!
\begin{array}{c||c||c|c||c|c|c||c|c|c|c||c|c|c|c|c||   }
\phantom{.}\j  &  0 & 1 && 2 &&& 3 &&&& 4 &&&&     \\
m \backslash i & 0 & 0 & 1 & 0 & 1 & 2 & 0 &1&2&3 & 0&1&2&3&4      \\
\hline
1  &  \underline{1,1} & \underline{Q,0} &  &&&&  &&&& &&&&   \\
2 &  1,0 & Q,-1 & \underline{\underline{\underline{1,-1}}}& \underline{\underline {1,0}}
& \underline{\underline{Q^2,-2}} & & \underline{Q^2, -1} &  &  &&&&  &&  \\
3 & 1,-1 & Q,-2 & 1,-2& 1,-1& Q^2,-3&\underline{\underline{\underline{\underline{1,-3}}}}
& Q^2,-2& \underline{\underline{\underline{1,-2}}}&\underline{\underline{\underline{Q^3,-4}}}
& & \underline{\underline{1,-1}}
&\underline{\underline{Q^3,-3}} &&&    \\
\hline
4 & 1,-2 & Q,-3 & 1,-3 & 1,-2& Q^2,-4& 1,-4 & Q^2,-3 & 1,-3 & Q^3,-5
& \underline{\underline{\underline{\underline{\underline{{ 1,-5}}}}}}
& 1,-2& Q^3,-4 & \underline{\underline{\underline{\underline{1,-4}}}}
&\underline{\underline{\underline{\underline{ { Q^4,-6}}}}} &    \\
5 & 1,-3 & Q,-4 & 1,-4& 1, -3& Q^2,-5& 1,-5 & Q^2,-4& 1,-4& Q^3,-6& { 1,-6}
& 1,-3 & Q^3,-5 &  1,-5 &Q^4,-7& { 1,-7} \\
6 & 1,-4 & Q,-5 & 1,-5& 1, -4& Q^2,-6& 1,-6 & Q^2,-5& 1,-5& Q^3,-7& 1,-7
& 1,-4 & Q^3,-6 &  1,-6 &Q^4,-8&  1,-8 \\
\ldots
\end{array} \nn
\ee
\be
\!\!\!\!\!\!\!\!\!\!\!\!\!\!\!\!\!\!\!\!\!\!\!\!\!\!\!\!\!\!\!\!\!\!\!\!
\begin{array}{c||c|c|c|c|c|c||c|c|c|c|c|c|c|| }
\phantom{.}\j  &  5 &&&&&& 6 &&&&&&     \\
m \backslash i & 0 & 1 & 2 & 3 & 4 & 5 & 0 &1&2&3 & 4&5&6      \\
\hline
1 & &&&&&&&&&&&&  \\
2 & &&&&&&&&&&&&   \\
3 & \underline{Q^3,-2} &&&&&&&& &&&&  \\
\hline
4 & Q^3,-3 &\underline{\underline{\underline{1,-3}}} &\underline{\underline{\underline{Q^4,-5}}}
&&&& \underline{\underline{1,-2}}
&  \underline{\underline{Q^4,-4 }}  &&&&&    \\
5 & Q^3,-4 &1,-4 &Q^4,-6
&\underline{\underline{\underline{\underline{\underline{{ 1,-6}}}}}}
&\underline{\underline{\underline{\underline{\underline{{ Q^5,-8}}}}}}&
& 1,-3 & Q^4,-5
&\underline{\underline{\underline{\underline{ { 1,-5}}}}}
&\underline{\underline{\underline{\underline{Q^5,-7}}}}&&&   \\
6 & Q^3,-5 &1,-5 & Q^4,-7 & 1,-7& Q^5,-9& 1,-9
& 1,-4& Q^4,-6& 1,-6& Q^5,-8& 1,-8 & Q^{6},-10 & \\
\ldots
\end{array} \ \ \ \ \ \nn
\ee
\be
\!\!\!\!\!\!\!\!\!\!\!\!\!\!\!\!\!\!\!\!\!\!\!\!
\begin{array}{c||c|c|c|c|c|c|c|c||c|c|c|c|c|c|c|c|c|c||   }
\phantom{.}\j  &  7 &&&&&&&& 8 &&&&&&&&&     \\
m \backslash i & 0 & 1 & 2 & 3 & 4 & 5 & 7 & 8 &  0 &1&2&3 & 4&5&6&7&8&9        \\
\hline
1 & &&&&&&&&&&&&&&&&& \\
2 & &&&&&&&&&&&&&&&&& \\
3 & &&&&&&&&&&&&&&&&&  \\
4 &   \underline{Q^4,-3} &&&&&&&&&&&&&&& &&    \\
5 &  Q^4,-4 &\underline{\underline{\underline{1,-4}}} & \underline{\underline{\underline{Q^5,-6}}}
&&&&&& \underline{\underline{1,-3}}&\underline{\underline{Q^5,-5}}&&&&&&& &  \\
6 &  Q^4,-5 & 1,-5 &Q^5,-7
&\underline{\underline{\underline{\underline{\underline{{ 1,-7}}}}}}
&\underline{\underline{\underline{\underline{\underline{{ Q^6,-9}}}}}}
 &&&& 1,-4&Q^5,-6&\underline{\underline{\underline{\underline{ { 1,-6}}}}}
&\underline{\underline{\underline{\underline{Q^6,-8}}}}&&&&&&   \\
\ldots
\end{array}\ \ \ \ \ \
\nn
\ee
\be
\begin{array}{c||c|c|c|c|c|c|c|c|c|c||c|c|c|c|c|c|c|c|c|c|c|c||c|c    }
\phantom{.}\j  &  9 &&&&&&&&&& 10 &&&&&&&&&&&& 11 &  \\
m \backslash i & 0 & 1 & 2 & 3 & 4 & 5 & 7 & 8 & 9 & 10 &  0 &1&2&3 & 4&5&6&7&8&9&10&11 &0& \ldots     \\
\hline
1 & &&&&&&&&&&&&&&&&&&&&&&&\\
2 & &&&&&&&&&&&&&&&&&&&&&&&\\
3 & &&&&&&&&&&&&&&&&&&&&&&& \\
4 & &&&&&&&&&&&&&&&&&&&&&&&   \\
5 &   \underline{Q^5,-4}&&&&&&&&&&&&&&&&&&&&&& \\
6 &   Q^5,-5&\underline{\underline{\underline{1,-5}}}&\underline{\underline{\underline{Q^6,-7}}}
&&&&&&&&\underline{\underline{1,-4}}&\underline{\underline{Q^6,-6}}&&&&&&&&&&
&\underline{Q^6,-5}  \\
\ldots
\end{array} \ \ \ \ \ \ \
\nn
\ee
}

Now we can use (\ref{w32m}) to restore the coefficients
(symmetrically deformed), and obtain
\be
W_3^{(2,m)}
= \sum_{j=0}^{2m-1} X_3^j \left(\sum_{i=0}^j
\frac{U_{i,j}\cdot  [m-1]_s!}{\left[{\rm floor}\left(\frac{j+i+1}{2}\right)\right]_s!
\left[m-{\rm floor}\left(\frac{j+i+2}{2}\right)\right]_s!}
\cdot Q^{{\rm deg}_{i,j}}[X_1+X_2]^{m+i}_{{\rm shift}_{i,j}}
X_1^{m-{\rm floor}\left(\frac{j+i+1}{2}\right)} X_2^{m-{\rm floor}\left(\frac{j+i+2}{2}\right)}
\right)
\label{W32m}
\ee
with shifts and degrees easily restored from the table.
Eq.(\ref{Omega32}) is considerably simpler after an additional reshuffling of terms
with different powers of $X_3$ and the use of $\Omega_2$ from (\ref{OmegaN2Q}):
\be
\Omega_3^{(2,m)} = \sum_{j=0}^m  \frac{[m]_q!}{[m-j]_q![j]_q!} \  X_3^j \cdot
\!\!\!
\underbrace{\left[X_3^2+X_1X_2\right]^{m-j}_{m-j+1}}_{\prod_{i=1}^{m-j} (X_3^2+Q^{m-j+1-2i}X_1X_2)}
\!\!\!\!\!\!
\cdot\ \Omega_3^{2,m+j}(X_1,X_2)
\ee
These $W$ and $\Omega$ can be substituted into (\ref{Z2vsW32}),
\be
\!\!\!\!\!\!  \Psi^{(2,m)}_{[0,1,0]}\! =\! [2m-1]_q Q^{m(m+1)} (Q^{-2m }X_2^2-X_3^2)\ W_{3}^{(2,m)}\!(X_1,X_2,X_3)
 +\! [3m-1]_q Q^{m(m-1)}\! X_3 \Omega_2^{(2,m)}\!(X_1,X_2,QX_3)
\label{Z2vsW32a}
\ee
and one can check that this is an $a=2$ twisted eigenfunctions.
It works.

\bigskip

Thus, now we have an algorithmic description of
$W_3^{(2,m)}$ for any $m$, starting from the $Q=1$ quantity $w_3^{(2,m)}(X_1,X_2,X_3)$.
The $Q=1$ formula (\ref{w32m}) at the origin of the algorithm is easy to obtain,
and it is easy to check that the result is an eigenfunction.
At the same time, solving the eigenfunction equations directly is difficult for
large values of $m$.
For other $N$ and $a$, the algorithm should be extended.

\subsection{Generic answers for the first level excitations at $N=3$
\label{N3genans}}

$Q=1$ expressions for the lowest states at $N=3$ are:
\be
\begin{array}{|c|c|c|}
{\rm Level}\ L=|\alpha| & {\rm excitation}\ \alpha &\psi_{\alpha} = \omega_3^{(a,m)}\times  \\
\hline
0 & [000] & 1 \\
1 & [001] &  X_3 \\
1 & [010] & (2m-1)X_2+ mX_3 \\
1 & [100] & (m-1)X_1+m(X_2 + X_3) \\
2 & [011] & X_2X_3 \\
2 & [101] & (2m-1)X_1X_3 + m X_2X_3 \\
2 & [110] & (m-1)X_1X_2 + m (X_1X_3+ X_2X_3) \\
2 & [002] & (m-1)X_3^2+  m(X_1X_3+ X_2X_3) \\
2 & [020] & 2(m-1)^2X_2^2+m(m-1)X_3^2 + 2m(m-1)X_1X_2 + m^2X_1X_3+m(3m-2)X_2X_3 \\
2 & [200] & (m-1)(m-2)X_1^2+m(m-1)(X_2^2+X_3^2) + 2m(m-1)(X_1X_2+X_1X_3) + 2m^2X_2X_3 \\
\ldots &&
\end{array}
\nn
\ee
The last two lines can be treated as linear combinations, implied by the previous one:
\be
2(m-1)\Big((m - 1)X_2^2 + mX_2(X_1+X_3)\Big) + m\Big((m - 1)X_3^2 + mX_3(X_1+X_2)\Big)
\ee
and
\be
(m - 2) \Big((m - 1)X_1^2 + mX_1(X_2+X_3)\Big)
+ m\Big((m - 1)X_2^2 + mX_2(X_1+X_3)\Big) + m\Big((m - 1)X_3^2 + mX_3(X_1+X_2)\Big)
\nn
\ee
but their deformation can be different.

\bigskip

The final complete answers for the three level one states at $N=3$ and $a=2$ are \cite{MMPns}:
\be
\Psi^{(2,m)}_{[001]} &=& X_3\cdot \Omega^{(2,m)}_{[001]}
\\ \nn \\
\Psi^{(2,m)}_{[010]} &=& [2m]_qX_3\,\Omega^{(2,m)}_{[001]} + 2 [2m-1]_qQ^{2m}X_2\,\Omega^{(2,m)}_{[010]}
+ \nn \\
&+& (Q^{2m}-1)[2m-1]_qX_2\Big\{Q^{m+1}X_3 \Big(X_1^2-Q^{-2(m-1)}X_2^2\Big)\,\Omega^{(2,m-1)}_{[011]} - \nn\\
&-& Q^{2m}X_1 \Big(X_2^2-Q^{-2(m-1)}X_3^2\Big)\,\Omega^{(2,m-1)}_{[001]}
+ Q^{3m}X_2\Big(X_3^2-Q^{-2m}X_1^2\Big)\,\Omega^{(2,m-1)}_{[000]}
\nn \\ \nn \\
\Psi^{(2,m)}_{[100]} &=& Q^{m-2\delta_{m,1}}  \frac{[2m]_q[m-1]_q}{[m]_q^2}X_1\,\Omega^{(2,m)}_{[011]}
+ 2Q^{2m}X_2\,\Omega^{(2,m)}_{[011]}+\nn\\
 &+& \frac{[2m]_q}{[m]_q}X_3\,\Omega^{(2,m)}_{[001]}
 + (Q^{2m}-1)X_2\Big\{Q^{m+1}X_3\Big(X_1^2-Q^{-2(m-1)}X_2^2\Big)\,\Omega^{(2,m-1)}_{[011]}-\nn \\
&-&Q^{2m}X_1\Big(X_2^2-Q^{-2(m-1)}X_3^2\Big)\,\Omega^{(2,m-1)}_{[001]}
+ Q^{3m}X_2\Big(X_3^2-Q^{-2m}X_1^2\Big)\,\Omega^{(2,m-1)}_{[000]}\Big\}\nn
\ee
where we introduced the notation $\Omega^{(a,m)}_{[a,b,c]}:=\Omega^{(a,m)}_3\Big(Q^aX_1,Q^bX_2,Q^cX_3\Big)$.

These formulas describe decomposition in $\Omega_3^{(2,m)}$ and $\Omega_3^{(2,m-1)}$,
which is not very practical beyond $(N,a) = (3,2)$ because these two quantities would differ by power
$3(a-1)$ in $X$, which is difficult to make from $a$-th powers like $X_i^a-X_j^a$.
Therefore generalizable are the $N=3$ formulas,
which do not involve $\Omega_3^{a,m-1}$ at all,
valid for all values of $a$ and $m$:
\be
\!\!\!\!\!\!\!\!
\boxed{\begin{array}{l}
\Psi^{(a,m)}_{[001]} = X_3\cdot \Omega^{(a,m)}_{[001]}
\nn \\ \nn \\
\Psi^{(a,m)}_{[010]} = [m]_qX_3\Omega^{(a,m)}_{[001]} + [2m-1]_qQ^{am}X_2\Omega^{(a,m)}_{[010]}
-(Q^a-1)[m]_q[2m-1]_qX_2^a
\frac{X_2\Omega^{(a,m)}_{[010]}-X_3\Omega^{(a,m)}_{[010]}}{X_2^a-X_3^a}
\sim \nn \\ \nn \\
\sim [2m-1]_q\cdot \frac{Q^{-am}X_2^a - X_3^a}{X_2^a-X_3^a}\cdot X_2\Omega^{(a,m)}_{[010]}
+ Q^{a(m-1)}[m]_q\cdot \frac{Q^{-a(2m-1)} X_3^a - X_2^a}{X_3^a-X_2^a} \cdot  X_3\Omega^{(a,m)}_{[001]}
\end{array}
}
\ee
As anticipated from simpler examples,
this answer, if expressed through $Q^a=q$ depends on $a$ only through $\Omega^{(a,m)}$.

It is noteworthy that
the relation in the second line is a polynomial, which is a characteristic feature of $\Omega$,
not following from either its symmetry
or from linear equations like (\ref{peri}).
However, this very ratio is not sufficient
for the third level one excitation at $N=3$.
A naive generalization of the above formula,
\be
\Psi^{(a,m)}_{[100]} \ \ \stackrel{a=1,2}{=}\
Q^{(a-1)m} [m-1]_qX_1\Omega^{(a,m)}_{[011]}
+Q^{am}[m]_qX_2\Omega^{(a,m)}_{[010]}  +[m]_qX_3\Omega^{(a,m)}_{[001]}
- \nn\\
- (Q^{am}-1)[m]_qX_2^a
\frac{X_2\Omega^{(a,m)}_{[010]}-X_3\Omega^{(a,m)}_{[001]} }{X_2^a-X_3^a}
\ee
works only at $a=1$ and $a=2$.
This is also reflected in the lack of universal dependence on $Q^a$.

The way out is to use a whole family of
{\it polynomial} quantities
\be
G^{(a,m)}_{i,j} := \frac{\hat T_i\Omega_3^{(a,m)}-\hat T_j\Omega_3^{(a,m)}}{X_i^a-X_j^a}, \nn \\
{\mathfrak G}^{(a,m)}_{i,j} := \frac{X_i^a \hat T_i\Omega_3^{(a,m)}-X_j^a \hat T_j\Omega_3^{(a,m)}}{X_i^a-X_j^a},
\nn \\
{\cal G}_1^{(a,m)} := \frac{\hat T_1\Omega_3^{(a,m)}}{(X_1^a-X_2^a)(X_1^a-X_3^a)}
+ \frac{\hat T_2\Omega_3^{(a,m)}}{(X_2^a-X_1^a)(X_2^a-X_3^a)} + \frac{\hat T_3\Omega_3^{(a,m)}}{(X_3^a-X_1^a)(X_3^a-X_2^a)}
\nn \\
{\cal G}_2^{(a,m)} := \frac{\hat T_2\hat T_3\Omega_3^{(a,m)}}{(X_1^a-X_2^a)(X_1^a-X_3^a)}
+ \frac{\hat T_1\hat T_3\Omega_3^{(a,m)}}{(X_2^a-X_1^a)(X_2^a-X_3^a)} + \frac{\hat T_1\hat T_2\Omega_3^{(a,m)}}{(X_3^a-X_1^a)(X_3^a-X_2^a)}
\label{polcom}
\ee
where we introduced a new operator $\hat T_i:=X_iQ^{X_i{\p\over\p X_i}}$.
Note that  $\hat T_3\hat T_3 \Omega_3^{(a,m)} = Q\ \Omega^{(a,m)}\Big(X_1,X_2,Q^2X_3\Big)$
produces an extra factor $Q$.

In these terms, we look for an eigenfunction in the form
\be
\Psi^{(a,m)}_{[100]} = v_1{\mathfrak G}^{(a,m)}_{2,3} + v_2{\mathfrak G}^{(a,m)}_{1,2}
+ (v_3X_1^a+v_4X_2^a+v_5X_3^a)G^{(a,m)}_{1,2} + (v_6X_2^a+v_7X_3^a)G^{(a,m)}_{2,3}
+ \nn \\
+ (v_8X_1^a+v_9X_3^a)G^{(a,m)}_{3,1}
+  v_{10}X_3^{2a} {\cal G}_1^{(a,m)}
\ee
where the coefficients $v_k$ are yet indeterminate.
They are, however, not fully fixed from the eigenvalue equation, and one possible
choice is
\be
\Psi^{(a,m)}_{[100]} \sim
Q^{a(m-1)}[m+1]_q{\mathfrak G}^{(a,m)}_{2,3} + (Q^{2am}+1)[m-1]_q{\mathfrak G}^{(a,m)}_{1,2}
+ \nn \\
+ \Big(-Q^{am}(Q^{am}+1)[m-1]_qX_2^a+ [m]_q^2(m-1)(Q^a-1)^2X_3^a\Big)G^{(a,m)}_{1,2}
- Q^{a(2m-1)} (X_2^a+[m]_qX_3^a)G^{(a,m)}_{2,3}
+ \nn \\
+Q^{am}[m-1]_q(-Q^{am}X_1^a+X_3^a)G^{(a,m)}_{3,1}
+  [m]_q^2[m-1[(Q^a-1)^2X_3^{2a} {\cal G}_1^{(a,m)}
\ee
We provide also the form preferred in \cite{MMPns}:
\be
\boxed{
\begin{array}{c}
\Psi^{(a,m)}_{[100]} =
Q^a[m-1]_q\cdot \frac{Q^{-am}X_1^a-X_2^a}{X_1^a-X_2^a}\cdot \frac{Q^{-am}X_1^a-X_3^a}{X_1^a-X_3^a}
\cdot\Omega^{(a,m)}_{[100]} + \nn \\ \nn \\
+[m]_q\cdot \frac{Q^{-a(m-1)}X_2^a-X_1^a}{X_2^a-X_1^a}\cdot \frac{Q^{-am}X_2^a-X_3^a}{X_2^a-X_3^a}
\cdot X_2\Omega^{(a,m)}_{[010]}
+ [m]_q\cdot \frac{Q^{-a(m-1)}X_3^a-X_1^a}{X_3^a-X_1^a}\cdot \frac{Q^{-am}X_3^a-X_2^a}{X_3^a-X_2^a} \cdot  X_3\Omega^{(a,m)}_{[001]}
\end{array}
}
\ee
Like the $Q\to 1$ case, this formula is explicitly symmetric in $X_2$ and $X_3$,
but not in $X_1$.

\bigskip

At the level two, there are six excitations.
Three of them are described by similar formulas:
\be
\boxed{
\begin{array}{c}
\Psi^{(a,m)}_{[011]} = X_2X_3\Omega^{(a,m)}_{[011]}  \\  \\
\!\!\!\! \Psi^{(a,m)}_{[101]} =
Q^{-a(m-1)} [2m-1]_q\cdot \frac{Q^{-am}X_1^a-X_2^a}{X_1^a-X_2^a}\cdot X_1X_3\Omega^{(a,m)}_{[101]}
+  \\
+ [m]_q\cdot \frac{Q^{-a(2m-1)}X_2^a-X_1^a}{X_2^a-X_1^a}\cdot X_2X_3\Omega^{(a,m)}_{[011]}
\\ \\
\Psi^{(a,m)}_{[110]} =
Q^{a } [m-1]_q \cdot\frac{Q^{-am}X_1^a-X_3^a}{X_1^a-X_3^a}\cdot\frac{Q^{-am}X_2^a-X_3^a}{X_2^a-X_3^a}
\cdot X_1X_2\Omega^{(a,m)}_{[110]} +  \\
+ [m]_q\left( \frac{Q^{-am}X_1^a-X_2^a}{X_1^a-X_2^a}\cdot\frac{Q^{-a(m-1)}X_3^a-X_2^a}{X_3^a-X_2^a}
\cdot X_1X_3\Omega^{(a,m)}_{[101]}
+\right. \\ \left.
+  \frac{Q^{-am}X_2^a-X_1^a}{X_2^a-X_1^a}\cdot \frac{Q^{-a(m-1)}X_3^a-X_1^a}{X_3^a-X_1^a}
\cdot X_2X_3\Omega^{(a,m)}_{[011]}\right)
\end{array}
}
\ee
The next excitation at the same level 2 is
\be
\ \ \boxed{
\begin{array}{c}
\Psi^{(a,m)}_{[002]} =
Q^{2a+1}[m-1]_q\cdot \frac{Q^{a(m-1)}X_2^a-X_3^a}{X_2^a-Q^aX_3^a}\cdot \frac{Q^{a(m-1)}X_1^a-X_3^a}{X_1^a-Q^aX_3^a}
\cdot X_3^2\Omega^{(a,m)}_{[002]} + \nn \\ \nn \\
+[m]_q\cdot \frac{Q^{am}X_2^a-X_1^a}{X_2^a-X_1^a}\cdot \frac{Q^{am}X_3^a-X_1^a}{Q^aX_3^a-X_1^a}
\cdot X_1X_3\Omega^{(a,m)}_{[101]}
+ [m]_q\cdot \frac{Q^{am}X_1^a-X_2^a}{X_1^a-X_2^a}\cdot \frac{Q^{am}X_3^a-X_2^a}{Q^aX_3^a-X_2^a} \cdot  X_2X_3\Omega^{(a,m)}_{[011]}
\end{array}
}
\ee
It involves a {\it new} polynomial(!) combination of rational functions, extending the set (\ref{polcom}).
In this particular case, denominators are obtained from looking at additional Q-shifts of ${\cal G}_1$
in (\ref{polcom}), and, say, ${\cal G}_2$ is not involved.

One can eliminate $Q$-powers from the above expressions
by additional manipulations, e.g.
\be
\Psi^{(a,m)}_{[002]} =
[m-1]_q\left(1+[m]_q(Q^a-1)\frac{X_1^a}{X_1^a-Q^aX_3^a}\right)\left(1+[m]_q(Q^a-1)\frac{X_2^a}{X_2^a-Q^aX_3^a}\right)
\hat T_3\hat T_3 \Omega_3^{(a,m)}
+ \nn \\
+ [m]_q\left(1+[m]_q(Q^a-1)\frac{X_2^a}{X_2^a- X_1^a}\right)\left(1+[m-1]_q(Q^a-1)\frac{Q^aX_3^a}{Q^aX_3^a-X_1^a}\right)
\hat T_1\hat T_3 \Omega_3^{(a,m)}
+ \nn \\
+ [m]_q\left(1+[m]_q(Q^a-1)\frac{X_1^a}{X_1^a-X_2^a}\right)\left(1+[m-1]_q(Q^a-1)\frac{Q^aX_3^a}{Q^aX_3^a-X_2^a}\right)
\hat T_2\hat T_3 \Omega_3^{(a,m)}
\ee
or
\be
\Psi^{(a,m)}_{[002]} =
\hat T_3\left\{[m-1]_q
\left(1+[m]_q(Q^a-1)\frac{X_1^a}{X_1^a- X_3^a}\right)\left(1+[m]_q(Q^a-1)\frac{X_2^a}{X_2^a- X_3^a}\right)
 \hat T_3 \Omega_3^{(a,m)} \right.
+ \nn \\
+ [m]_q\left(1+[m]_q(Q^a-1)\frac{X_2^a}{X_2^a- X_1^a}\right)\left(1+[m-1]_q(Q^a-1)\frac{ X_3^a}{ X_3^a-X_1^a}\right)
\hat T_1\hat  \Omega_3^{(a,m)}
+ \nn \\ \left.
+ [m]_q\left(1+[m]_q(Q^a-1)\frac{X_1^a}{X_1^a-X_2^a}\right)\left(1+[m-1]_q(Q^a-1)\frac{ X_3^a}{X_3^a-X_2^a}\right)
\hat T_2\hat  \Omega_3^{(a,m)}\right\}
:= \hat T_3 {\cal K}^{(a,m)}_{1,2,3}
\ee

All eigenfunctions we discussed above can be represented in the same way:
\be
\Psi^{(a,m)}_{[001]} = \hat T_3\Omega_3^{(a,m)}
\label{Z001fin}
\ee
\be
\Psi^{(a,m)}_{[010]} =
[2m-1]_q\left(1+[m]_q(Q^a-1)\frac{X_3^a}{X_3^a-X_2^a}\right)\hat T_2 \Omega_2^{(a,m)}
+ [m]_q\left(1+[2m-1]_q(Q^a-1)\frac{X_2^a}{X_2^a-X_3^a}\right)\hat T_3\Omega_3^{(a,m)}
= \nn
\ee
\be
= [2m-1]_q\left(1+[m]_q\cdot \frac{q-1}{1-\frac{x_2 }{x_3 }}\right)\hat T_2 \Omega_2^{(a,m)}
+ [m]_q\left(1+[2m-1]_q\cdot \frac{q-1}{1-\frac{x_3}{x_2}}\right)\hat T_3\Omega_3^{(a,m)}
\ee
\be
\Psi^{(a,m)}_{[100]} =
[m-1]_q\left(1+[m]_q(Q^a-1)\frac{X_2^a}{X_2^a-X_1^a}\right)\left(1+[m]_q(Q^a-1)\frac{X_3^a}{X_3^a-X_1^a}\right)
\hat T_1\Omega_3^{(a,m)} + \nn\\
+ [m]_q\left(1+[m-1]_q(Q^a-1)\frac{X_1^a}{X_1^a-X_2^a}\right)\left(1+[m]_q(Q^a-1)\frac{X_3^a}{X_3^a-X_2^a}\right)
\hat T_2\Omega_3^{(a,m)} + \nn\\
+ [m]_q\left(1+[m-1]_q(Q^a-1)\frac{X_1^a}{X_1^a-X_3^a}\right)\left(1+[m]_q(Q^a-1)\frac{X_2^a}{X_2^a-X_3^a}\right)
\hat T_3\Omega_3^{(a,m)}
\ee
\be
\Psi^{(a,m)}_{[011]} = \hat T_2\hat T_3\Omega_3^{(a,m)}
\ee
\be
\Psi^{(a,m)}_{[101]} = \hat T_3\left\{
[2m-1]_q\left(1+[m]_q(Q^a-1)\frac{X_2^a}{X_2^a-X_1^a}\right)\hat T_1\Omega_3^{(a,m)}
+ [m]_q\left(1+[2m-1]_q(Q^a-1)\frac{X_1^a}{X_1^a-X_2^a}\right)\hat T_2\Omega_3^{(a,m)}
\right\}
\nn
\ee
\be
\Psi^{(a,m)}_{[110]} =
[m-1]_q\left(1+[m]_q(Q^a-1)\frac{X_3^a}{X_3^a-X_1^a}\right)\left(1+[m]_q(Q^a-1)\frac{X_3^a}{X_3^a-X_2^a}\right)
\hat T_1\hat T_2\Omega_3^{(a,m)} + \nn \\
+ [m]_q\left(1+[m]_q(Q^a-1)\frac{X_2^a}{X_2^a-X_1^a}\right)\left(1+[m-1]_q(Q^a-1)\frac{X_2^a}{X_2^a-X_3^a}\right)
\hat T_1\hat T_3\Omega_3^{(a,m)} + \nn \\
+ [m]_q\left(1+[m]_q(Q^a-1)\frac{X_1^a}{X_1^a-X_2^a}\right)\left(1+[m-1]_q(Q^a-1)\frac{X_1^a}{X_1^a-X_3^a}\right)
\hat T_2\hat T_3\Omega_3^{(a,m)}
\ee
\be
\Psi^{(a,m)}_{[002]} =
\hat T_3\left\{[m-1]_q
\left(1+[m]_q(Q^a-1)\frac{X_1^a}{X_1^a- X_3^a}\right)\left(1+[m]_q(Q^a-1)\frac{X_2^a}{X_2^a- X_3^a}\right)
 \hat T_3 \Omega_3^{(a,m)} \right.
+ \nn \\
+ [m]_q\left(1+[m]_q(Q^a-1)\frac{X_2^a}{X_2^a- X_1^a}\right)\left(1+[m-1]_q(Q^a-1)\frac{ X_3^a}{ X_3^a-X_1^a}\right)
\hat T_1\hat  \Omega_3^{(a,m)}
+ \nn \\ \left.
+ [m]_q\left(1+[m]_q(Q^a-1)\frac{X_1^a}{X_1^a-X_2^a}\right)\left(1+[m-1]_q(Q^a-1)\frac{ X_3^a}{X_3^a-X_2^a}\right)
\hat T_2\hat  \Omega_3^{(a,m)}\right\}
\label{Z002fin}
\ee
An advantage of such a representation is that one can search for the rational functions $\frac{X_i^a}{X_i^a-X_j^a}$,
which are the only non-trivial part of the formula, in the first order in $\eta:=Q^a-1=q-1$,
and restore additional factors $Q^a$ in front of $X_i^a$ afterwards.
Actually, such factors are absent in (\ref{Z001fin})-(\ref{Z002fin}), which can thus be considered
as ideal examples of the ``proper'' deformation.

If one treats in these formulas $\eta$ as a new independent variable,
then {\bf at each power of $\eta$, the expressions are polynomials in $X$ (!)}.
Of course, they are eigenfunctions only at the particular value $\eta=Q^a-1$,
thus this is a statement concerning the space of polynomial combinations of $\Omega$,
but this is interesting subject by itself.
Using $g_n(x)$ from (\ref{g}),
one can rewrite these formulas in a compact form, alternative to the presentation in \cite{MMPns}
and providing a {\it proper} deformation of the table at the beginning of s.\ref{N3genans}
(it consists of an artful insertion of the $g$-factors
and of the placement of the operators $\hat T_i$):
{\footnotesize
\be
\!\!\!\!\!\!\!\!\!\!\!\!\!\!\!\!\!\!\!\!\!\!\!\!\!\!\!\!
\boxed{
\begin{array}{c}
\Psi^{(a,m)}_{[001]} = \hat T_3\Omega_3^{(a,m)}
\\ \\
\Psi^{(a,m)}_{[010]}
= [2m-1]_q\cdot g_m\!\left(\frac{x_2}{x_3}\right)\cdot \hat T_2 \Omega_3^{(a,m)}
+ [m]_q\cdot g_{2m-1}\!\left(\frac{x_3}{x_2}\right)\cdot \hat T_3 \Omega_3^{(a,m)}
\\ \\
\Psi^{(a,m)}_{[100]}
= [m-1]_q\cdot g_m\!\left(\frac{x_1}{x_2}\right)g_m\!\left(\frac{x_1}{x_3}\right)\cdot \hat T_1 \Omega_3^{(a,m)}
+[m]_q\cdot g_m\!\left(\frac{x_2}{x_1}\right)g_m\!\left(\frac{x_2}{x_3}\right)\cdot \hat T_2 \Omega_3^{(a,m)}
+[m]_q\cdot g_m\!\left(\frac{x_3}{x_1}\right)g_m\!\left(\frac{x_3}{x_2}\right)\cdot \hat T_3 \Omega_3^{(a,m)}
\\ \\
\Psi^{(a,m)}_{[011]} = \hat T_2\hat T_3\Omega_3^{(a,m)}
\\ \\
\Psi^{(a,m)}_{[101]} = \hat T_3\left\{
[2m-1]_q\cdot g_m\!\left(\frac{x_1}{x_2}\right)\cdot \hat T_1 \Omega_3^{(a,m)}
+ [m]_q\cdot g_{2m-1}\!\left(\frac{x_2}{x_1}\right)\cdot \hat T_2 \Omega_3^{(a,m)}
\right\}
\\ \\
\Psi^{(a,m)}_{[110]} =
[m-1]_q\cdot g_m\!\left(\frac{x_1}{x_3}\right)g_m\!\left(\frac{x_2}{x_3}\right)
\cdot \hat T_1\hat T_2 \Omega_3^{(a,m)}
+[m]_q\cdot g_m\!\left(\frac{x_1}{x_2}\right)g_{m-1}\!\left(\frac{x_3}{x_2}\right)
\cdot \hat T_1\hat T_3 \Omega_3^{(a,m)}
+[m]_q\cdot g_m\!\left(\frac{x_2}{x_1}\right)g_{m-1}\!\left(\frac{x_3}{x_1}\right)
\cdot \hat T_2\hat T_3 \Omega_3^{(a,m)}
\\ \\
\Psi^{(a,m)}_{[002]} =
\hat T_3\left\{
[m-1]_q\cdot g_m\!\left(\frac{x_3}{x_1}\right)g_m\!\left(\frac{x_3}{x_2}\right)\cdot \hat T_3 \Omega_3^{(a,m)}
+[m]_q\cdot g_m\!\left(\frac{x_1}{x_2}\right)g_{m-1}\!\left(\frac{x_1}{x_3}\right)\cdot \hat T_1 \Omega_3^{(a,m)}
+[m]_q\cdot g_m\!\left(\frac{x_2}{x_1}\right)g_{m-1}\!\left(\frac{x_2}{x_3}\right)\cdot \hat T_2 \Omega_3^{(a,m)}
\right\}
\\ \\
\!\Psi^{(a,m)}_{[020]} \!\!= \!
[2m-2]_q g_m\!\left(\frac{x_2}{x_3}\right)
\cdot\hat T_2\left\{
[m-1]_qg_m\!\left(\frac{x_2}{x_1}\right)g_m\!\left(\frac{x_2}{x_3}\right)\hat T_2 \Omega_3^{(a,m)}
\!+ [m]_qg_{m-1}\!\left(\frac{x_1}{x_2}\right)g_m\!\left(\frac{x_1}{x_3}\right)\hat T_1 \Omega_3^{(a,m)}
\!+ [m]_qg_{m-1}\!\left(\frac{x_3}{x_2}\right)g_m\!\left(\frac{x_3}{x_1}\right)\hat T_3 \Omega_3^{(a,m)}
\right\}  + \!\!\! \\
+   [m]_q g_{2m-2}\left(\frac{x_3}{x_2}\right)
\cdot\hat T_3\left\{
[m-1]_qg_m\!\left(\frac{x_3}{x_1}\right)g_m\!\left(\frac{x_3}{x_2}\right)\hat T_3 \Omega_3^{(a,m)}
+ [m]_qg_{m}\!\left(\frac{x_1}{x_2}\right)g_{m-1}\!\left(\frac{x_1}{x_3}\right)\hat T_1 \Omega_3^{(a,m)}
+ [m]_qg_{m}\!\left(\frac{x_2}{x_1}\right)g_{m-1}\!\left(\frac{x_2}{x_3}\right)\hat T_2 \Omega_3^{(a,m)}
\right\}
\\ \\
\!\!\!\Psi^{(a,m)}_{[200]} \!\!= \!
[m-2]_q g_m\!\left(\frac{x_1}{x_2}\right)\!g_m\!\left(\frac{x_1}{x_3}\right)
\!\hat T_1\!\!\left\{
[m-1]_qg_m\!\left(\frac{x_1}{x_2}\right)\!g_m\!\left(\frac{x_1}{x_3}\right)\hat T_1 \Omega_3^{(a,m)}
\!\!+\! [m]_qg_{m-1}\!\left(\frac{x_2}{x_1}\right)\!g_m\!\left(\frac{x_2}{x_3}\right)\hat T_2 \Omega_3^{(a,m)}
\!\!+\! [m]_qg_{m-1}\!\left(\frac{x_3}{x_1}\right)\!g_m\!\left(\frac{x_3}{x_2}\right)\hat T_3 \Omega_3^{(a,m)}
\right\}  + \!\!\! \\
\!+[m]_q g_{m-2}\!\left(\frac{x_2}{x_1}\right)g_m\!\left(\frac{x_2}{x_3}\right)
\cdot\!\hat T_2\!\left\{
[m-1]_qg_{m-2}\!\left(\frac{x_2}{x_1}\right)g_m\!\left(\frac{x_2}{x_3}\right)\hat T_2 \Omega_3^{(a,m)}
\!+ [m]_qg_{m-1}\!\left(\frac{x_1}{x_2}\right)g_m\!\left(\frac{x_1}{x_3}\right)\hat T_1 \Omega_3^{(a,m)}
\!+ [m]_qg_{m-1}\!\left(\frac{x_3}{x_2}\right)g_m\!\left(\frac{x_3}{x_1}\right)\hat T_3 \Omega_3^{(a,m)}
\right\}  + \!\!\! \\
+[m]_q g_{m-2}\!\left(\frac{x_3}{x_1}\right)g_m\!\left(\frac{x_3}{x_2}\right)
\cdot\hat T_3\left\{
[m-1]_qg_m\!\left(\frac{x_3}{x_1}\right)g_m\!\left(\frac{x_3}{x_2}\right)\hat T_3 \Omega_3^{(a,m)}
\!+ [m]_qg_{m-1}\!\left(\frac{x_1}{x_3}\right)g_m\!\left(\frac{x_1}{x_2}\right)\hat T_1 \Omega_3^{(a,m)}
\!+ [m]_qg_{m-1}\!\left(\frac{x_2}{x_3}\right)g_m\!\left(\frac{x_2}{x_1}\right)\hat T_2 \Omega_3^{(a,m)}
\right\}
\\ \\
\ldots
\end{array}
}
\nn
\ee
}
One can already observe certain structure and regularity in these formulas
but we refrain from making precise statements at this stage.

\section{Conclusion}

In this paper, we describe eigenfunctions $\Psi_{\alpha}^{(a,m)}$ of the twisted Cherednik integrable system.
We begin with the limit $q\longrightarrow 1$, when they are much simpler.
At this level, it is easy to observe a hierarchical structure with
a symmetric polynomial of peculiar variables $X_i=x_i^{1/a}$ as the ground state $\omega_N^{(a,m)}$
and its non-symmetric descendants (non-symmetric Jack polynomials) at higher levels,
enumerated by \wcs, non-ordered integer partitions of the level $L=|\alpha|$.
Remarkably, excitations do not depend on $a$.

At $q \neq 1$, the pattern remains just the same, with symmetric $\Omega_N^{(a,m)}[\vec X]$ in the background
and non-symmetric $a$-independent excitations.
However, their extraction from eigenvalue equations becomes a technically complicated problem.
One way to construct
the symmetric polynomial   $\Omega_N^{(a,m)}[\vec X]$  is to solve the Chalykh periodicity equations
\cite{Cha,MMPCha}, (\ref{peri}) in any two variables, and then to use an additional property of symmetricity.
Symmetric solutions of a given grading are then unique, and the full $\Omega_N^{(a,m)}[\vec X]$
is restored.

Despite this is a significant simplification, the problem remains computationally difficult and still unsolvable
in this way for big enough $N,a,m$. Hence, we discussed in this paper another approach, constructing some ``deformation prescription''
by a {\it direct} $q$-deformation of formulas from the $q\to 1$ limit.
An ideal example is provided by (\ref{OmegaN2Q}) at $N=2$,
where the only non-trivial step is division of a single factor $m$ into two differently deformed pieces,
$m-k$ and $k$, consistent with the structure of the formula,
which looks almost combinatorial.
Then we try to extend the same trick to other $m$.
Partly the attempt is already successful, partly not quite:
we stop at the point where alarming coefficients appear in two boxes in (\ref{Omega333}).
This can imply the need for splitting like $m= m-k+k$  in (\ref{OmegaN2Q}),
and we leave this story for further investigation.

Also we did not elaborate too much on the story of excitations:
this is because it opens a whole new chapter in the theory of integrable systems,
revealing the need for non-symmetric polynomials,
which so far appeared not too much within this context.
We devoted a separate big text \cite{MMPns} to this new development,
which has overlaps with the present paper, but puts accents differently
and attracts attention to somewhat different aspects of emerging theory.

Concerning concrete issues that look more technical than conceptual,
the next steps could include:
\begin{itemize}
\item{\bf Ground state.}
Generic answers for the ground states $\Omega_N^{(a,m)}$.
In this and the preceding papers, they are found  for
\be
\begin{array}{c|c|c|c}
N & a & m & \\
\hline
2 & \forall & \forall & (\ref{34}) \\
3 & 2 & \forall &  (\ref{Omega32})   \\
  &  3  & 1,2,3,4,5,6  & (\ref{67}) \ \& \ \hbox{Ome3.txt} \\
4 & 2 & 1,2 & (\ref{41})  \\
5 & 2 & 1  & (\ref{43})
\end{array}
\ee
This is already something, but not enough. This paper contains explanation and ideas about the derivations and possible condensed notation. As to explicit resulting formulas/answers, except for just a few, they are somewhat lengthy,
and are collected in the file Ome3.txt, attached to this submission.

\item{\bf Excitations.}
Generic answers for the excitations.
In \cite{MMPns} and here, they were found only for levels $L=1,2$,
it remains to be done.

\item{\bf Creation operators.}
The fact of principal importance is that the  answers for excitations
are functorial in $\Omega$, i.e. look like $a$-independent universal operators,
applied to a given ground state.
This means that one can search for the answer without knowing explicit
expressions for $\Omega$.
However, the formulas {\it per se} are applicable only at correct $\Omega$,
otherwise they are not even polynomial in $X$-variables.
Thus there is a conspiracy between the creation operator, which provides
the excitation and the ground state.
In particular, the creation operators are not yet found in the ``free'' form,
when they can multiply by each other (applied several times).
At the moment, the situation is more complicated than just the lack of
commutativity, observed already in \cite{KN,MMcreops}.
Not only different creation operators are mutually dependent,
they also depend on the ground state.
Still it looks like they do exist(!).
Then they seem to form a rich algebra, which, in many respects,
resembles an algebra of representations
(a category standing at a would-be other side of the Tannaka-Krein
duality to DIM commutation relations) with non-trivial braiding
and decomposition properties of the products.
This is all the more supported by appearance of cut-and-join like
formulas \eqref{eq:cut-n-join-like-rec}.

\item{\bf Polynomiality puzzle.}
It looks like there are plenty of rational operators
(with differences of $X$ or, probably, of $x=X^a$ in denominators),
which produce {\it polynomials} when acting on a peculiar polynomial $\Omega$.
This fact does not look like an immediate corollary of either eigenvalue, or Chalykh
equations, and can imply some {\it additional} properties of $\Omega$,
additional to the fact that they are the only eigenfunctions
which are {\it symmetric} polynomials, and then (and only then) they simultaneously
are twisted Baker-Akhiezer functions.
There is still something more in these symmetric polynomials.
The first step in this direction could be just classification of allowed
rational operators, a few are mentioned in (\ref{polcom}), many more are
provided by $\eta$ expansion at the end of the previous section.
\end{itemize}

To conclude, we proposed an approach in this paper to look for complicated solutions of eigenvalue equations
in the form of deformation of simple
solutions of the their $q\to 1$ (Jack case) counterparts.
The point is to {\bf find a {\it form} of the $q\to 1$ solutions,
where the deformation is {\it direct}} (we also call it ``proper''):
ordinary integers are just replaced by quantum numbers,
powers of sums of variables are replaced by Pochhammer-like products,
and additional powers of $q$ are inserted.
Given a {\it family} of quantities, one can identify the
``proper'' basis and realization of above three rules by looking at the first
members of the family, and then {\it use} them for everything else.
Once known, the answer can be easily validated.
As usual for NP-like problems, {\bf it is difficult to find an answer,
but it is easy to check it}.

However, NP problems are non-typical for fundamental physics,
and the resolution in this case comes through the concept
``quasiclassics is exact'', which is typical for non-perturbative physics
and seems to work also in the case of twisted Cherednik operators.
Ideal from the point of view of this paper are the formulas (\ref{Z001fin})-(\ref{Z002fin})
and those in the following table,
which express the excitations through the vacuum state $\Omega$,
these expressions can be {\bf directly deformed} from the $q\to 1$ limit.
This adds to the mounting evidence of applicability of Duistermaat-Heckman (DH) like
approach to non-perturbative physics \cite{Niemi,UFN2,UFN3}.
It is also known to be natural for integrable systems, and we see now that it can be true
for their highly non-trivial generalizations too \cite{MMMP2,MMP}
based on associative algebras rather than on Lie algebras.
It would be very interesting to continue and develop this approach
and extend examples worked out in \cite{MMPns} and in the present paper
to a full-fledged and rigorous theory.

At this stage, exact deformation remains a piece of art rather than deductive science,
and this paper adds one more portion of evidence that one should look for a better formulation.
Usually such formulations are provided by hidden action of global groups
(like those in the DH theorem) or by hidden nilpotent symmetries
(like those in the index theorems and their homological or supersymmetric reformulations).
An adequate formulation directly extendable to the DIM algebra
is still lacking, and should be worked out.

\section*{Acknowledgements}

This work is supported by the RSF grant 24-12-00178.

\end{document}